\documentclass[a4paper,11pt]{article}
\pdfoutput=1
\usepackage{jcappub} 
\usepackage[T1]{fontenc}
\usepackage[titletoc,title]{appendix}
\newcommand{\beq}{\begin{equation}}
\newcommand{\eeq}{\end{equation}}
\usepackage{graphicx}
\graphicspath{{./Figures/}}
\usepackage{amsmath}
\usepackage{MnSymbol}
\usepackage{relsize}
\usepackage{mathrsfs}
\usepackage{subcaption}
\usepackage{bigints}
\usepackage{gensymb}
\title{\boldmath Closing the window on $\sim$GeV Dark Matter with moderate ($\sim$$\mu$b) interaction with nucleons}
\author{M. Shafi Mahdawi}
\author{and Glennys R. Farrar}
\affiliation{Center for Cosmology and Particle Physics, Department of Physics, New York University, \\4 Washington Place, New York, NY 10003, USA}
\emailAdd{shafi.mahdawi@nyu.edu}
\emailAdd{gf25@nyu.edu}

\abstract{We improve limits on the spin-independent scattering cross section of Dark Matter on nucleons, for DM in the 300 MeV -- 100 GeV mass range, based on the DAMIC and XQC experiments.  Our results close the window which previously existed in 1 -- 8 GeV mass range, for a DM-nucleon cross section of order $\sim \mu$b, assuming the standard velocity distribution.}

\begin{document}
\maketitle
\flushbottom
\section{Introduction}
As constraints on Weakly Interacting Massive Particle (WIMP) DM have become more and more stringent, and most of the parameter space for WIMP DM compatible with LHC bounds has been ruled out, interest in DM in the mass range below 10 GeV has increased;  see~\cite{Gelmini2016} for a recent review.   
\par
The existence of a window for moderately-interacting DM ($\sigma_p \sim \mu$b where $\sigma_p$ is DM-nucleon cross section) in the 0.3 -- 100 GeV mass range was emphasized long ago~\cite{ZF_window}.  Since that time both the lower and upper limits of the allowed cross section regime have moved.  First, the re-analysis of XQC by Erickcek et al.~\cite{Erickcek2007} concluded that the upper edge of the allowed region was actually higher than determined by~\cite{ZF_window}, such that a window remained in spite of the constraint from below based on DAMIC~\cite{Kouvaris2014}, which superseded the limit from CRESST~\cite{CRESST2002} used in~\cite{ZF_window}. Here, we re-examine the limits from XQC and DAMIC pertinent to the existence of a window for $\sigma_p \sim \mu$b. We find in fact the window is closed by these experiments.  
\par
XQC --- the X-ray Quantum Calorimeter~\cite{McCammon2002} --- was an X-ray detector aboard a sounding-rocket.  Due to its small overburden it was sensitive to DM with relatively strong interactions.  The XQC data was first used to place a rough upper bound on the DM-proton cross section in~\cite{Wandelt2000}, then the bound was refined in~\cite{ZF_window}.  Later, ~\cite{Erickcek2007} claimed the bound of ~\cite{ZF_window} to be overly stringent. Checking the analysis in~\cite{Erickcek2007}, caused us to suspect that those authors assumed that the XQC detector is shielded by the body of the sounding-rocket;  they have confirmed this was indeed their assumption (A. Erickcek private communication). However for the lower boundary of the $\sigma_p$ region that can be excluded by the XQC, the DM has negligible interaction probability in the rocket body\footnote{We only examine here the ``window region'' at $\sigma_p \sim \mu$b and have not considered other boundaries of the excluded region reported by Erickcek et al., which were also not examined in~\cite{ZF_window}.}.  Thus our result for the low-cross-section reach of XQC is almost a factor 10 stronger than that of~\cite{Erickcek2007}. While our result updates and tightens in some mass regions the bounds of~\cite{ZF_window}, it is largely consistent with the original analysis in~\cite{ZF_window}.  
\par
The DAMIC --- Dark Matter In CCDs~\cite{Barreto2012264}  --- experiment was operated at a depth of $=$106.7 meters underground.  As a result, DAMIC is shielded both by the Earth and the lead shield around it. If the DM-nucleon cross section is large enough, the DM particles lose too much energy in collisions  prior to reaching the detector, to be able to make an energy deposit above threshold for the detector\footnote{We denote as \emph{capable}, those particles reaching the detector with enough energy to potentially trigger it.}. Thus DAMIC only places a constraint on DM, for cross sections \emph{below} some limit. Prior to the work reported here, the only analysis of the DAMIC data to address the moderately-interacting window was that of Kouvaris and Shoemaker~\cite{Kouvaris2014}, KS below, who used a mean scattering-length and energy-loss approximation, and assumed purely vertical propagation, following Starkman et al.~\cite{Starkman}, SGED below. 
\par
We report here the results of our detailed Monte-Carlo simulation of DM particles propagating in the Earth's crust to reach DAMIC's depth. We developed an importance sampling technique to make the otherwise very numerically intensive simulations tractable; details of this method are given in~\cite{method}. We find that as much as four orders-of-magnitude more events actually would be seen, for the limiting cross section of the analysis found in~\cite{Kouvaris2014} using the SGED\textbf{/}KS approximation, than observed by DAMIC (see figure~\ref{DAMIC_att@sigma_SGED} and section~\ref{SGED}). This discrepancy translates to 1.8 -- 5.6 times stronger limit on the DM-proton cross section from DAMIC depending on the DM mass. The DM{\scriptsize ATIS} code, which we developed for this study, will be publicly available on-line~\cite{code}.
\par
Our improved limits from reanalyzing the XQC and DAMIC data, close the window for moderately-interacting DM in the mass range 1 -- 8 GeV\footnote{Rich et al.~\cite{RICH1987} reports exclusion of $\mu$b DM candidate heavier than 2.1 GeV. Unfortunately, a detailed description of their silicon detector, exact time of the flight, exposure time, location and orientation of the detector is not provided in~\cite{RICH1987}, so we cannot evaluate that analysis. To be more conservative, we restrict to limits using XQC and DAMIC. }, assuming the standard DM velocity distribution.  We note the strong dependence of the upper bound on the DM-nucleon cross section implied by the XQC data on the DM velocity distribution. We show how the limit depends on the local DM velocity distribution, for some representative DM masses. 
\section{Preliminaries}
The number of DM particles with velocities in the interval $(\vec{v},\vec{v}+d^3v)$, which pass through a surface element $d\vec{A}$ of the target in a detector during a time interval $dt$, is
\beq\begin{split}
	\label{e_NDM}
	\frac{dN_{DM}}{d^3v\,dt}=\frac{\rho}{m}\,f(\vec{v},\vec{v}_{det})\,\vec{v}\cdot d\vec{A}=\frac{\rho}{m}\,v\,f(\vec{v},\vec{v}_{det})\,cos\,\theta\, dA \;,
\end{split}\eeq
where $\rho$ is the DM local mass density, $m$ is the mass of each DM particle, $f(\vec{v},\vec{v}_{det})$ is the unit-normalized velocity distribution of DM particles in the detector rest frame, and $\theta$ is the angle between the DM velocity vector and the normal vector to the surface element.
\par
The probability of a DM particle scattering off a nucleus of mass number $A$ in the path length element $dl$ of the target along the trajectory of this DM particle, denoted by $dP_A$, is
\beq\begin{split}
	\label{e_PA}
	dP_A=&\frac{dl}{\lambda_A}=\frac{1}{\lambda_A}\,\frac{dz}{cos\,\theta}\;,
\end{split}
\eeq
where $dz$ is the projection of the path length element $dl$ along the normal vector of the target's surface element. $\lambda_A\equiv(n_A\,\sigma_{A})^{-1}$ is the mean interaction length of the DM particle passing through a target of nuclear mass number $A$ and number density $n_A$, for DM-nucleus cross section $\sigma_A$.
\par
Using eqs.~\eqref{e_NDM} and~\eqref{e_PA} and allowing for a mixed composition target, the number of nuclei in a target of volume $V$ scattered by DM particles with velocities in $(\vec{v},\vec{v}+d^3v)$ during the time interval $dt$ in the limit of single scattering ($P_A\ll1$) is
\beq\begin{split}
	\label{e_Ndv}
	\frac{dN}{d^3v\,dt}=&\int_V\frac{dN_{DM}}{d^3v\,dt}\,\sum_{A}dP_A\\=&\frac{\rho}{m}\,v\,f(\vec{v},\vec{v}_{det})\sum_{A} \frac{M_A\,\sigma_A}{m_A}\;,
\end{split}\eeq
where $M_A/m_A=n_AV=N_A$ is the number of nuclei of mass number $A$ in the target. Eq.\eqref{e_Ndv} shows that only the total mass $M_A$ of nuclei each of mass number $A$ in the target determines the number of scattered nuclei in each bin of the DM's velocity. 
\par
A DM particle with speed $v$ and energy $E=\frac{1}{2}\,m\,v^2$ in the detector rest frame transfers part of its initial energy to the target nucleus in the detector. This nuclear recoil energy is
\beq
\begin{split}
	\label{e_Er}
	E_{r}&=\frac{4\,\mu^2_{A}}{m\,m_A}\,E\,\left(\frac{1-\cos\,\xi^{CM}}{2} \right)\\
	&=\hspace{3mm}\frac{\mu^2_{A}}{m_A}\hspace{2mm}v^2\,\left(\,1-\cos\,\xi^{CM} \,\right)\;,
\end{split}
\eeq
where $\xi^{CM}$ is the scattering angle in the CM frame and $\mu_A$ is the DM-nucleus reduced mass. 
\par 
The momentum transfer is small for a light ($m<100$ GeV) DM particle that scatters off a nucleus. Therefore, the corresponding wavelength is larger than the nuclear radius. This has two consequences: 
\begin{itemize}
	\item The scatterings in the CM frame are isotropic.
	\item The spin-independent scattering amplitudes of individual nucleons add up coherently:
	\beq
	\label{e_MatrixElement}
	\mathcal{M}^{SI}\propto(Zf_p+(A-Z)f_n)\;,
	\eeq	
	where $f_n$ and $f_p$ are the neutron and proton spin-independent amplitudes; $A$ and $Z$ are the mass number and the atomic number respectively.
\end{itemize}
\par
In a general discussion, Kurylov and Kamionkowski~\cite{Kurylov2004} showed that in the small momentum transfer limit only the scalar and the axial vector terms of the interaction Lagrangian survive. Using this result, the DM-proton spin-independent cross section is
\beq\begin{split}
	\label{e_sigmaN}
	\sigma_p&=\frac{4\,f_p^2}{\pi}\mu_p^2\;.
\end{split}\eeq
Specializing to the case of equal neutron and proton spin-independent couplings in eq.~\eqref{e_MatrixElement}, the spin-independent DM-nucleus cross section in the small momentum transfer limit is
\beq\begin{split}
	\label{e_sigmaA}
	\sigma_{A}^{0}&=\sigma_p\, \left(\frac{\mu_A}{\mu_p} \right) ^2\,A^2   \;.
\end{split}\eeq
\par
In real collisions with non-zero momentum transfer, not all DM-nucleon scattering amplitudes add up coherently. When the momentum transfer, defined by $q\equiv\sqrt{2m_AE_{r}}$, increases such that the corresponding wavelength is no longer large compared to the nuclear radius, the cross section begins to fall with the increase of $q$. This effect can be taken into account by a form factor, $F(qr_A)$, where $r_A$ is the effective nuclear radius. The effective nuclear radius $r_A$ can be approximately found by fitting the muon scattering data to a Fermi distribution~\cite{fricke1995}
\beq
\label{e_NuclearRadius}
r_A^2=c^2+\frac{7}{3}\pi^2a^2-5s^2\;,
\eeq
with parameters: $c\simeq(1.23A^{1/3}-0.6)$ fm, $a\simeq0.52$ fm, and $s=0.9$ fm. 
We use the following analytical expression which is proposed by Helm~\cite{Helm} to calculate the nuclear form factor
\beq
\label{e_FormFactor}
F_A(E_r)=F(qr_A)=3\,\frac{\sin(qr_A)-qr_A\cos(qr_A)}{(qr_A)^3}\,e^{-(qs)^2/2}\;.
\eeq
\par
The DM-nucleus cross section in the non-zero momentum transfer limit and the DM-proton cross section in the zero momentum transfer limit are related by:
\beq\begin{split}
	\label{e_sigmaAF2}
	\sigma_{A}&=\sigma_p \left(\frac{\mu_A}{\mu_p} \right) ^2A^2\,F_A^2(E_r)   \;.
\end{split}\eeq
\par
Due to the isotropy of the scatterings in the CM frame, DM particles with a given speed produce a uniform distribution of recoil energies, i.e. $E_r\in\mathcal{U}(0, \,2\,\mu^2_{A}v^2/m_A)$. So, the differential energy-deposit rate is~\cite{Lewin199687}
\beq\begin{split}
	\label{e_NdEr}
	\frac{dN}{dE_r\,dt}=&\int_{v_{min}(E_r,A)}\frac{dN}{d^3v\,dt}\,\frac{m_A}{2\,\mu^2_A\,v^2}\,d^3v\\=&\frac{\rho}{2\,m}\sum_{A}\int_{v_{min}(E_r,A)} \frac{M_A\,\sigma_A}{\mu^2_A}\frac{f(\vec{v},\vec{v}_{det})}{v}\,d^3v\;,
\end{split}\eeq
where the lower limit of integration, $v_{min}$, is the minimum speed that a DM particle needs in order to deposit the recoil energy $E_{r}$ into the detector:
\beq
\label{e_vm}
v_{min}(E_r,A)=\sqrt{\frac{m_A\,E_{r}}{2\,\mu^2_{A}}}\;.
\eeq
\par
Substituting eq.~\eqref{e_sigmaAF2} in eq.~\eqref{e_NdEr}, the differential detection rate in time interval $dt$ is
\beq\begin{split}
	\label{e2_NdEr}
	\frac{dN}{dE_r\,dt}=&\frac{\rho\,\sigma_p}{2\,m\,\mu^2_p}\sum_{A}M_A\,A^2F_A^2(E_r)\int_{v_{min}(E_r,A)} \frac{f(\vec{v},\vec{v}_{det})}{v}\,d^3v\;.
\end{split}\eeq

\section{XQC bounds}\label{XQC}
To find the upper bound on the DM-proton cross section for $\mu$b DM, we use the observed energy-deposit spectrum of the X-ray Quantum Calorimetry (XQC) experiment, a rocket experiment launched on March 28, 1999. The Earth shielded the XQC detector from part of the DM flux. The body of the rocket carrying the XQC detector might also shield the detector, depending on the DM interaction length. In the following subsections, we will consider both of these shielding scenarios to calculate the expected energy-deposit spectrum of XQC and the corresponding DM-proton cross section upper bound.
 \subsection{Velocity distribution and geometrical shielding factors}
\begin{figure}[tbp]\centering
	\begin{subfigure}{.5\textwidth}
		\centering
		\includegraphics[width=\linewidth]{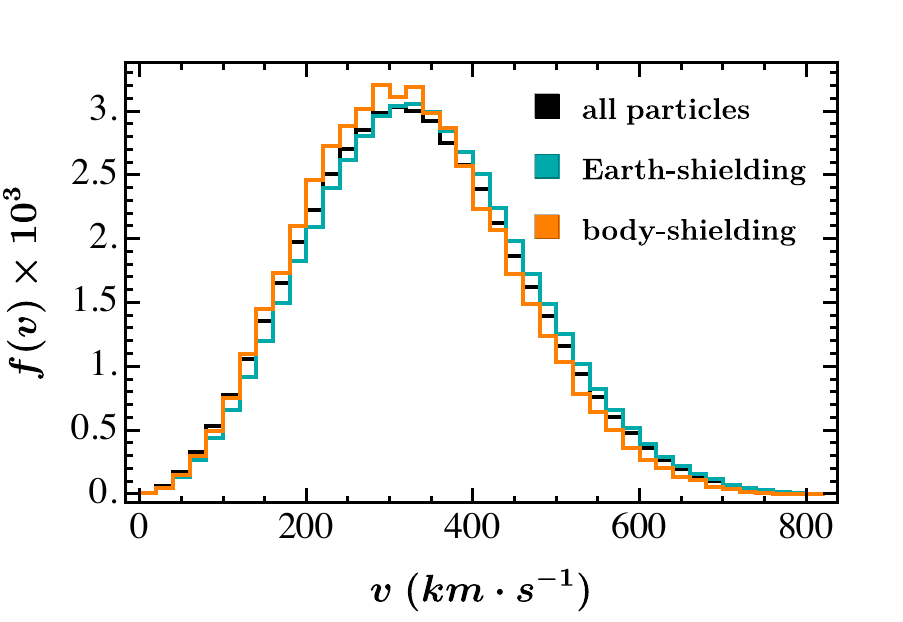}
		\caption{}
		\label{v_norm_dis}
	\end{subfigure}
	\begin{subfigure}{.49\textwidth}
		\centering
		\includegraphics[width=\linewidth]{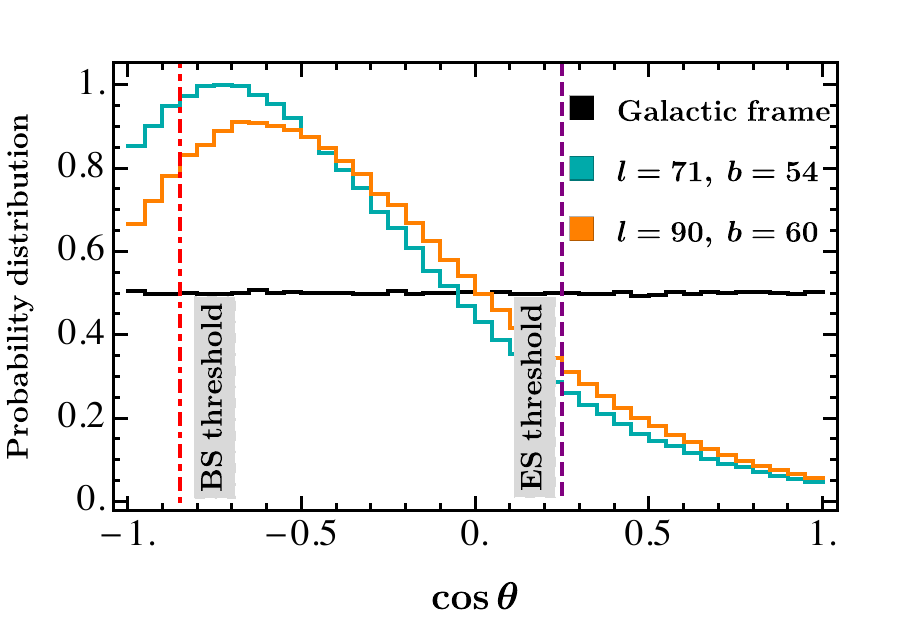}
		\caption{}
		\label{v_angle_dis}
	\end{subfigure}
	\caption{The truncated Maxwellian velocity distribution at the Earth, for $v_0=220\,\rm{km\cdot s^{-1}}$, $v_{esc}=584\,\rm{km\cdot s^{-1}}$, and $\vec{v}_{E}=(39.14,\,230.5,\,3.57) \,\rm{km\cdot s^{-1}}$. (a) The unit-normalized speed distribution of all DM particles (black), DM particles that are not shielded by Earth (cyan), and DM particles that are not shielded by rocket body (orange). (b) The zenith angle distribution of DM particles in the Galactic rest frame (black), relative to the Earth at flight launch (cyan), and relative to the XQC apparatus at detection time (orange). A DM particle hitting the detector from above, moving antiparallel to the $\hat{z}$ axis, has $cos\,\theta=$ -1 in the relevant coordinate system (see figure~\ref{XQC_par}).  The purple dashed line corresponds to the maximum zenith angle of a DM particle able to reach XQC in the Earth-shielding scenario while the red dash-dotted line shows the corresponding maximum zenith angle in the body-of-rocket shielding scenario. XQC's alignment serendipitously almost maximized the DM flux.}
	\label{f_Maxwellian}
\end{figure}

 Assuming an isothermal spherical density profile, the DM velocity distribution in the Galactic rest frame, $f_G(\vec{v})$, is characterized by a Maxwellian distribution with a velocity dispersion of $v_0$ truncated at an escape velocity $v_{esc}$. Correspondingly, the velocity of DM particles in the detector frame can be calculated by subtracting the velocity of the detector in the Galactic rest frame, $\vec{v}_{det}$, from the DM velocity in the Galactic rest frame $\vec{v}_G$. The black histogram in figure \ref{v_norm_dis} shows the DM speed distribution seen by a detector that moves with the Earth's velocity, $\vec{v}_{E}=(39.14,\,230.5,\,3.57) \,\rm{km\cdot s^{-1}}$, in the Galactic rest frame and Galactic coordinate system, for velocity dispersion $v_0=220\,\rm{km\cdot s^{-1}}$ and escape velocity $v_{esc}=584\,\rm{km\cdot s^{-1}}$~\cite{Erickcek2007}. 
 \par
\begin{figure}[tbp]\centering
	\includegraphics[width=0.65\textwidth]{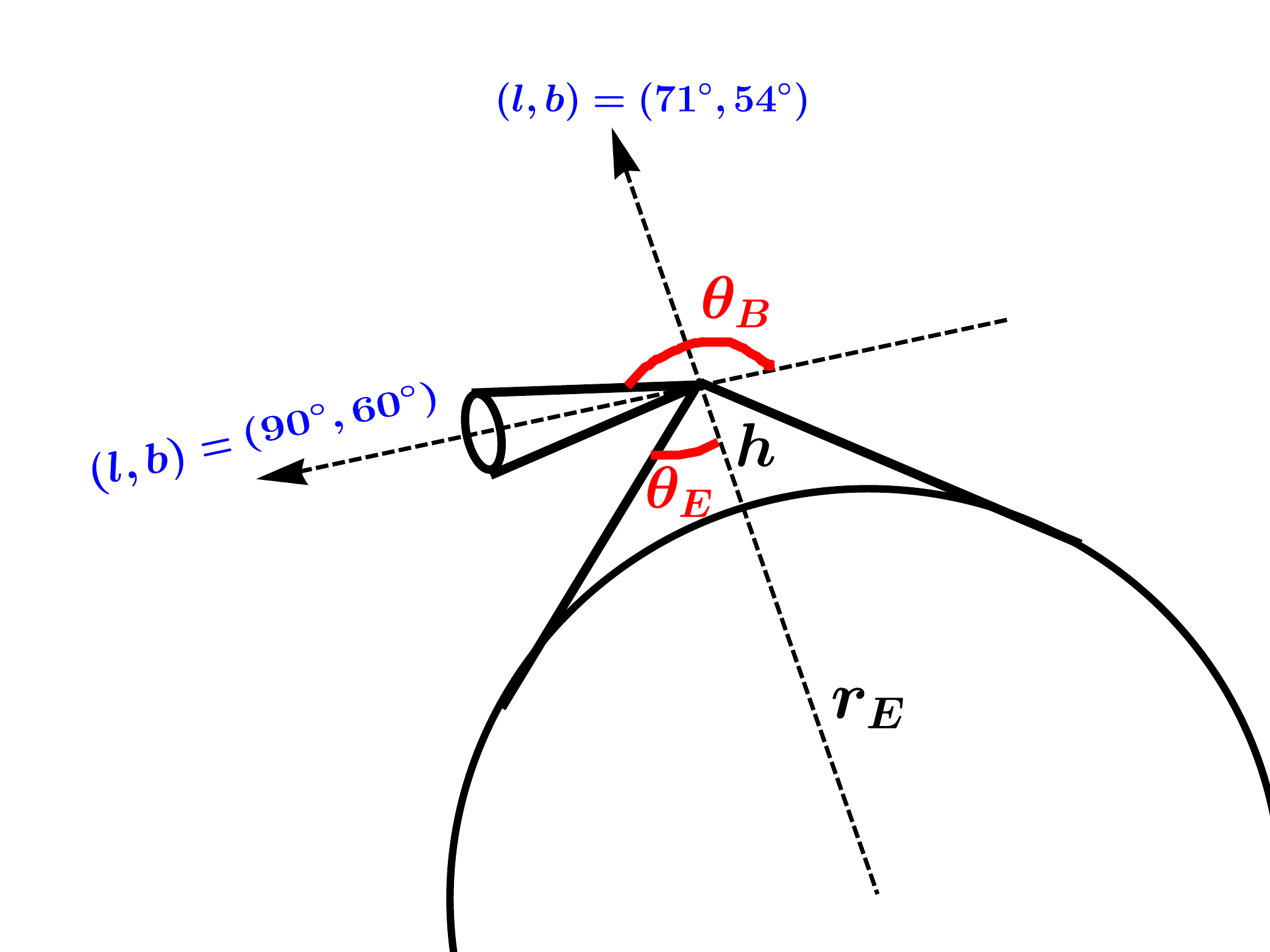}
	\caption{\label{XQC_par} Schematic drawing of the two coordinate systems pertinent to XQC. Dashed lines correspond to the $\hat{z}$ axis of each coordinate system. Zenith angle zero corresponds to a DM particle that reaches the detector from below along the $\hat{z}$ axis in each coordinate system.}
\end{figure}

XQC flight 27.041UG was launched 1999 March 28 at 09:00UT from White Sands, New Mexico~\cite{McCammon2002}. The normal vector to the Earth's surface at the position and time of flight launch was in the direction $(l,b)=(71\degree, 54\degree)$ in the Galactic coordinate system. The data was collected a few minutes after flight launch at a time-averaged height of 201 km above the Earth's surface, when the normal vector of the detector was pointing toward $(l,b)=(90\degree, 60\degree)$. Figure \ref{v_angle_dis} shows the zenith angle distribution of DM particles in various coordinate systems centered on the XQC detector. $cos\,\theta=$ -1 corresponds to a DM particle that reaches the detector from above along the $\hat{z}$ axis in each coordinate system. The cyan histogram shows the zenith angle distribution of DM particles in the coordinate system whose $\hat{z}$ axis is normal to the Earth's surface at the position and time of flight launch. The orange histogram is the zenith angle distribution in the coordinate system whose $\hat{z}$ axis is aligned with the normal vector to the surface of the XQC detector during the detection period. Figure~\ref{XQC_par} shows a schematic view of these two coordinate system and their corresponding $\hat{z}$ axis. 
\par 
\begin {table}[t] 
\begin{center}	
	\begin{tabular}{||c|c|c|c||} 
		\hline
		i& $E_r^i\,(eV)$& $\mathrm{f_e}(E_r)$ & $O_i $\\ [0.1ex] 
		\hline\hline
		1& 29-36& 0.38 & 0\\ 
		\hline
		2& 36-128& 0.5 & 11\\ 
		\hline
		3& 128-300&1&129 \\
		\hline
		4& 300-540&1&80\\
		\hline
		5& 540-700&1&90\\ 
		\hline
		6& 700-800&1&32\\ 
		\hline
		7& 800-945&1&48\\
		\hline
		8& 945-1100&1&31\\
		\hline
		9& 1100-1310&1&30\\
		\hline
		10& 1310-1500&1&29\\
		\hline
		11& 1500-1810&1&32\\
		\hline
		12& 1810-2505&1&15\\
		\hline
		13& $\geq$4000&1&60\\
		\hline
	\end{tabular}
\end{center} 
\caption {The XQC sensitivity factor and the observed number of events in each recoil energy bin. Bin 2505-4000 eV is ignored due to its contamination by the detector's interior calibration sources. }\label{t1}
\end {table}
In the case that the body of the rocket is essentially transparent to DM particles, only the Earth shields the XQC detector. The horizontal displacement of the rocket is negligible compared to its vertical displacement and the Earth's radius. To find the geometrical shielding factor in this case, we use the velocity distribution in the frame with the $\hat{z}$ axis pointing toward the flight zenith, $(l,b)=(71\degree, 54\degree)$. Using the time-averaged height of the XQC measurement, $h\,\rm{=201}$ km, DM particles must have at least the following zenith angle to be able to reach the detector
\beq
\label{e_thetasE}
\theta_E=sin^{-1} \left( \frac{r_E}{r_E+h}\right)\,\approx75.8\degree \;,
\eeq
where $r_E$ is the Earth's radius. The purple dashed line in figure \ref{v_angle_dis} corresponds to this angle. The geometrical shielding factor in this case is
\beq
\label{e_fsE}
K_E=N_{\theta}^{ -1}\int^{\theta_E}_{0}f(\vec{v},\vec{v}_{l})\,sin\,\theta\,d\theta\,\approx0.098\;,
\eeq
where $f(\vec{v},\vec{v}_{l})$ is the DM velocity distribution in the frame with its normal axis pointing toward vertical direction at the launch of the rocket carrying XQC and $N_{\theta}=\int_{0}^{\pi}f(\vec{v},\vec{v}_l)\,sin\theta\,d\theta$ is the normalization factor.
\par
If the body of the rocket carrying XQC completely shields the detector, only DM particles coming through the 1 steradian opening angle centered on the normal to the detector, $(l,b)=(90\degree, 60\degree)$, would reach the detector; this was the implicit assumption of Erickcek et al.~\cite{Erickcek2007} (private communication). The red dash-dotted line at cos$\,\theta_B\approx-0.85$ in figure \ref{v_angle_dis} corresponds to the minimum zenith angle of a DM particle that reaches the XQC detector. The Earth was not blocking any part of the 1 steradian opening angle as expected since the purpose of the mission was to detect astrophysical diffuse X-ray sources. The geometrical shielding factor in the body-shielding scenario is
\beq
\label{e_fsb}
K_B=N_{\theta}^{ -1}\int^{\theta_B}_{0}f(\vec{v},\vec{v}_{det}) \,sin\,\theta\,d\theta\,\approx0.891\;.
\eeq  

\subsection{Cross section limits}
Each of the 34 XQC calorimeters was composed of a 0.96 $\mu$m film of HgTe mounted on a 14 $\mu$m substrate of Si ~\cite{McCammon2002}. The total mass of each component of the XQC detector is
\beq
\label{e12}
M_A=\begin{cases}
	1.30\times10^{-3}\,g\hspace{1cm} A=Si\\
	1.02\times10^{-4}\,g  \hspace{1cm} A=Te\\
	1.61\times10^{-4}\,g \hspace{1cm} A=Hg\;.
\end{cases}
\eeq
The exposure time of the experiment depended on the recoil energy threshold, $E_r$, so we write $t_e(E_r)=100.7\,\mathrm{f_e}(E_r)$ seconds, where $\mathrm{f_e}(E_r)$ is the fraction of the exposure time that XQC was sensitive to recoil energy $E_r$. Table \ref{t1} contains the sensitivity factors and the observed number of events in each bin taken from Erickcek et al.~\cite{Erickcek2007}.
\begin{figure}[tbp]\centering
	\includegraphics[width=1\textwidth]{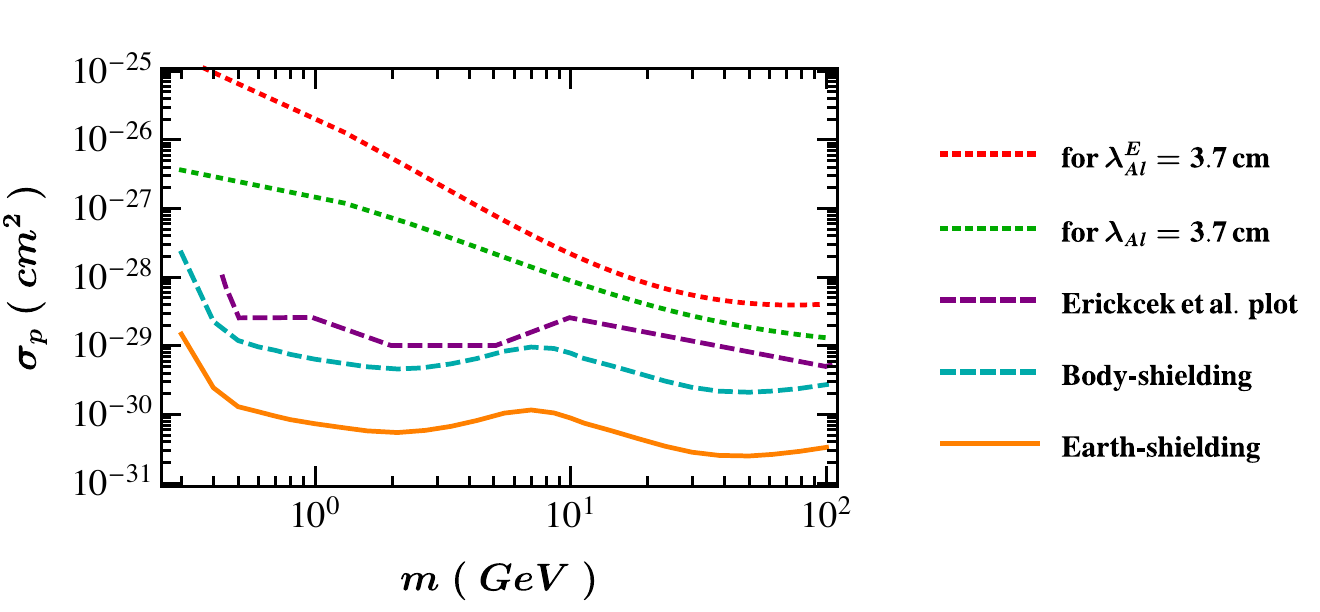}
	\caption{\label{XQC_bounds}The XQC 90\% CL upper bounds assuming the body of the rocket (dashed cyan) or the Earth (solid orange) shields the XQC detector. The dashed purple line is read from Erickcek et al. paper~\cite{Erickcek2007}. The green and red dotted lines show the DM-proton cross section required for a 3.7 cm interaction length and energy-loss length respectively in aluminum. Since the XQC overburden is few centimeters equivalent of aluminum, the rocket body does not shield the XQC detector.}
\end{figure}
\par
Substituting the geometrical shielding factors introduced in the previous subsection into eq.~\eqref{e2_NdEr}, the XQC differential detection rate for each shielding scenario, labeled by ${\{E,B\}}$ for the Earth-shielding and body-shielding scenarios respectively, is
\beq\begin{split}
	\label{eXQC_NdEr}
	\frac{dN_{\{E,B\}}}{dE_r}=&\,100.7\,\mathrm{f_e}(E_r)\,(1-K_{\{E,B\}})\,\frac{\rho\,\sigma_p}{2\,m\,\mu^2_p}\sum_{A}M_A\,A^2F_A^2(E_r)\int_{v_{min}(E_r,A)}v\,f_{\{E,B\}}(v)\,dv\;,
\end{split}\eeq
where $f_E(v)$ and $f_B(v)$ are the unit-normalized speed distributions of DM particles which are not shielded by the Earth and rocket body respectively (refer to figure~\ref{v_norm_dis}).
\par
The differential detection rate can be calculated for a given DM mass, $m$, a trial DM-proton cross section, $\sigma_p$, and taking the local DM density $\rho=$ 0.3 $\rm{GeV cm^{-3}}$. The crudest approach is to pick an interval from the observed energy-deposit spectrum whose signal to expected background ratio is the most sensitive and find the 90\% CL DM-proton cross section. This method was used by Zaharijas and Farrar~\cite{ZF_window} to analyze the XQC data. Erickcek et al.~\cite{Erickcek2007} used a more sophisticated approach which exploits the shape of the observed energy-deposit spectrum by binning this spectrum and calculating a figure of merit to find the 90\% CL DM-proton cross section upper bound. Yellin~\cite{Yellin1, Yellin2} argues that still a better method to find the  cross section upper bound for experiments with an uncertain background is to use the maximum gap method or the optimal interval method depending on the number of events in a deposit-energy interval. In the following, we adopt the Erickcek et al. method, because the Yellin method requires individual event energies to which we do not have access. While the Erickcek et al. method may not be optimal according to~\cite{Yellin1, Yellin2}, using it enables us to directly compare our result with theirs. Moreover, given that we close the window by increasing the DAMIC exclusion regime, having the most stringent possible limit from XQC is not essential.
\par
The basic strategy of Erickcek et al.~\cite{Erickcek2007} is to step through trial choices of cross section and for each of them define a quantity
\beq
\label{eChi2}
\chi_\sigma^2\equiv\sum_{i=1}^{13}\frac{(N_i(\sigma)-O_i)^2}{N_i(\sigma)}~\Theta(N_i(\sigma) - O_i)\;,
\eeq
where $O_i$ is the observed number of events and $N_i(\sigma)$ is the expected number of events for the given trial cross section in the i-th energy bin, and $\Theta$ is the Heaviside function. Only bins in which the predicted number is larger than the observed number are included in the sum. Thus as the trial cross section is decreased from some large value, both the $N_i$ values decrease and also bins which contribute at larger $\sigma$ become excluded from the sum. Therefore, $\chi_\sigma^2$ is a monotonically decreasing function of $\sigma$, eventually becoming zero when $N_i(\sigma) \leq O_i$ in every bin. The 90\% CL upper limit on the cross section is defined to be the value of $\sigma$ for which 90\% of Poisson-distributed datasets for the expected $N_i$ values, have a $\chi_\sigma^2$ value smaller than in the data.
Specifically:
\begin{itemize}
	\item from Poisson's distribution with predicted mean value $N_i$, the observed number of events in each bin is sampled to find the mock $O_i$'s.
	\item The $\chi^2$ value is calculated for each mock $\{O_i\}$, using eq.~\eqref{eChi2}.
	\item The number of cases in which the value of $\chi^2$ is smaller than the original value of $\chi^2$ is counted, and the confidence level is calculated using CL $=$ counts/(total number of trials).
\end{itemize}
The inferred DM-proton cross section limit is taken to be the trial cross section for which this fraction is 90\% CL.
\par
Figure~\ref{XQC_bounds} shows the DM-proton cross section upper bound as a function of the DM mass for the two different shielding scenarios. The body-shielding bounds are bigger than the bounds in the Earth-shielding scenario by a factor of 8.9 on average. The difference is primarily due to the ratio of the transmission factors, i.e. $\frac{1-K_E}{1-K_B}\approx8.3$, with a small difference due to the slightly different speed distributions of DM particles in the Earth-shielding and the body-shielding scenarios (refer to figure~\ref{v_norm_dis}).
\par
The aluminum body of the rocket carrying XQC produces an overburden of $\approx10\,\rm{g\cdot cm^{-2}}$ (\cite{McCammon2002} and Erickcek private communication~\cite{ErickcekThesis}), corresponding to 3.7 cm aluminum. The green dotted line in figure~\ref{XQC_bounds} shows the DM-proton cross section corresponding to the DM interaction length for an aluminum target of 3.7 cm thickness. The red dotted line shows the cross-section corresponding a mean energy-loss length of 3.7 cm in aluminum. The mean energy-loss length is
\beq
\label{elambdaE}
\lambda_{A}^E\equiv\left(E^{-1}\frac{\left\langle E_{r,A}\right\rangle}{\lambda_A} \right)^{-1} \;,
\eeq 
where $\left\langle E_{r,A}\right\rangle\equiv\frac{2\,\mu_A^2}{m\,m_A}\,E$ is the average nuclear recoil energy of a DM particle scattering off a nucleus of mass number $A$. 
\par
For a $m=$100 GeV DM particle, the probability of interacting with an aluminum nucleus in the rocket body is $\approx 2\%$ (in figure~\ref{XQC_bounds}, for $m=$100 GeV, compare the value of solid orange line, 3.3$\times\rm{10^{-31}\,cm^2}$, and the value of green dotted line, 1.5$\times\rm{10^{-29}\,cm^2}$). Therefore even for the heaviest DM candidate considered here, the body of the XQC rocket needs to be at least 50 times denser or thicker than 3.7 cm aluminum to begin to shield the detector significantly. We conclude that the rocket body did not shield the XQC detector for the smallest values of $\sigma_p$ that can be ruled out by the XQC experiment. Thus the appropriate scenario is the only-Earth-shielding one.
\begin{figure}[tbp]\centering
	\includegraphics[width=0.75\textwidth]{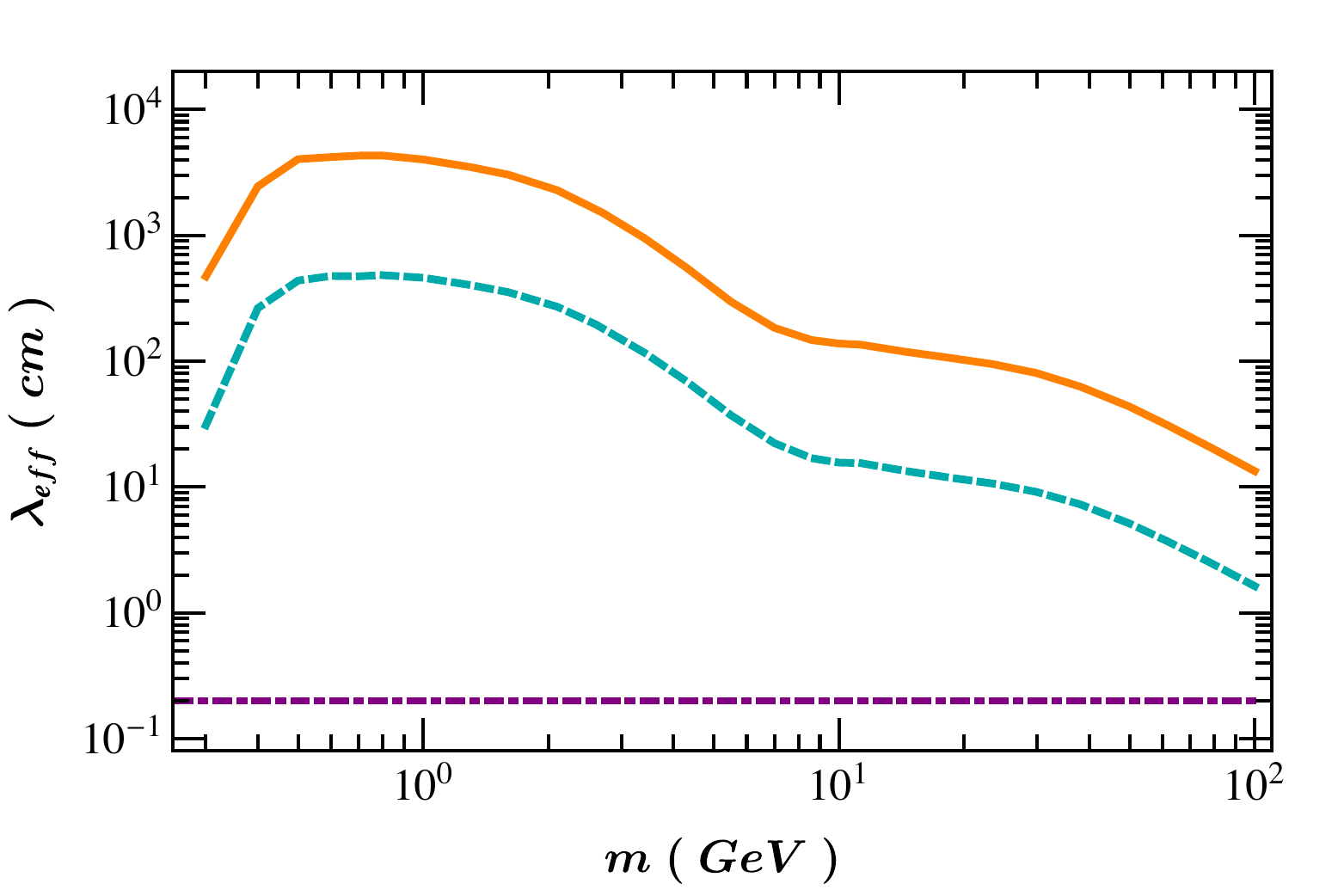}
	\caption{\label{XQC_lambdas} The mean interaction lengths of a DM particle in the XQC detector --- made of Si, Te, and Hg --- using the  cross section bound from the Earth-shielding (solid orange) and body-shielding (dashed cyan). The dashed purple horizontal line shows the typical size of the XQC detector; this legitimizes the sufficiency of the single-scattering approximation in calculating the deposit-energy rate in eq.~\eqref{eXQC_NdEr}.}
\end{figure}
\par
Finally, we check the validity of our simplification in ignoring the geometry of the XQC detector itself and using the total mass of each type of nuclei in the XQC detector to calculate the deposit-energy rate in eq.~\eqref{eXQC_NdEr}. The mean interaction length of DM particles in a target with a mix of nuclei can be calculated by
\beq
\label{eLambdaeff}
\lambda_{eff}^{-1}=\sum_{A}\lambda_{A}^{-1}=\sum_{A}n_{A}\,\sigma_A\;,
\eeq 
where $A\in\{Si,Te,Hg\}$ for the XQC detector. Figure \ref{XQC_lambdas} shows this mean interaction length using the DM-proton cross section upper bound in the two shielding scenarios. The mean interaction length in the body-shielding and the Earth-shielding scenarios are respectively at least 10 and 100 times bigger than the typical size of the XQC detector ($\approx 2$ mm). Thus DM particles do not have multiple scatterings in the XQC detector for the smallest values of $\sigma_p$ that are ruled out by the XQC experiment and eq.~\eqref{eXQC_NdEr} is accurate.
\subsection{Sensitivity to the DM velocity distribution}
\begin{figure}[tbp]\centering
	\includegraphics[width=.95\textwidth]{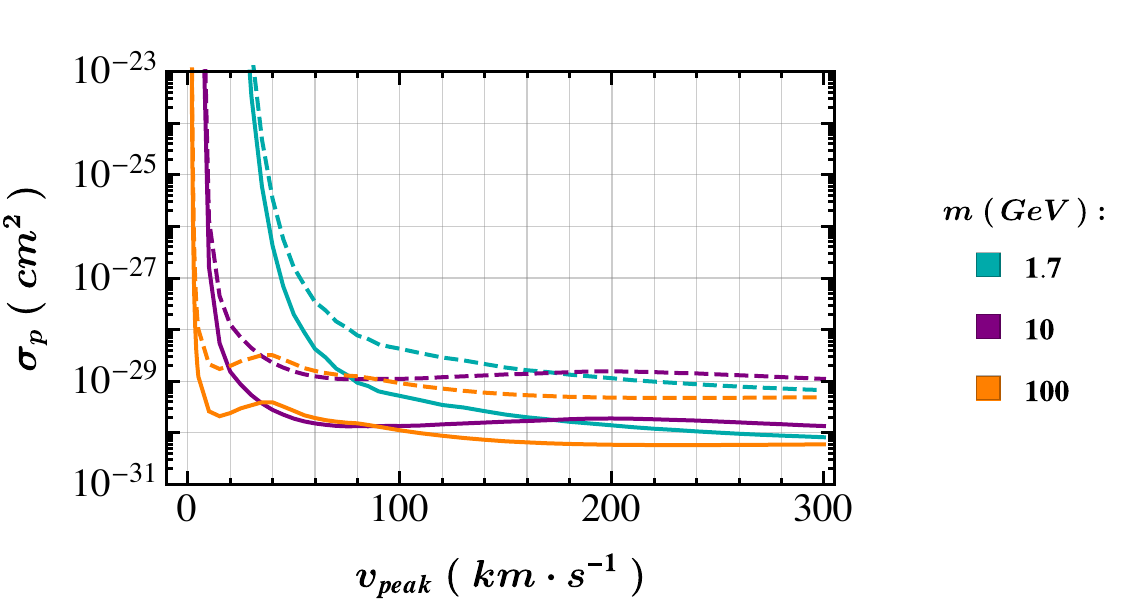}
	\caption{\label{XQC-vpeak}XQC's 90\% CL upper bound on the DM-proton cross section, as a function of the peak velocity of a Maxwellian distribution which characterizes the speed distribution of DM particles in the XQC rest frame, for three representative DM masses. Solid (dashed) lines correspond to the Earth (body) shielding scenario.  Although for large peak velocities the strongest bound is governed by the Earth-shielding scenario, the body-shielding scenario is applicable for cross sections above $\approx 0.5 \times 10^{-27} {\rm cm}^2$.}
\end{figure}

In the above discussions, we have only considered the usual DM velocity distribution. However, as discussed in~\cite{gfinprep17}, for a large enough cross section, DM interactions with gas in the Galactic disk can in principle significantly alter the local DM velocity distribution, causing the DM to co-rotate with the solar system to some degree. Given that the energy available to deposit in XQC depends quadratically on the velocity, reducing the DM relative velocity distribution reduces the flux of particles capable of producing an energy deposit above threshold.  We illustrate this effect by determining the cross section bounds obtained from XQC, under the assumption that the DM velocity relative to XQC is given by a Maxwellian with peak velocity $v_{peak}$.  Figure~\ref{XQC-vpeak} shows how, as the relative speed of DM particles decreases, the upper bound on DM-proton cross section increases and XQC's bounds become weaker.  Note that for cross-sections above $\approx 0.5 \times 10^{-27}$, the body-shielding limit applies.  For example, if 1.7 GeV DM particles co-rotate with the baryonic matter such that their speed distribution is a truncated Maxwellian in the XQC rest frame a window opens for peak velocity $v_{peak}\leq 95\,\rm{km\cdot s^{-1}}$, so that both the XQC and DAMIC bounds can be evaded.  
\section{DAMIC bounds}
With the weakly interacting massive particle (WIMP) DM paradigm in mind, most DM direct detection experiments are built underground. The layer of the Earth's crust acts as a shield that filters out unwanted cosmic-ray and radioactive backgrounds while not affecting the DM flux. For a $\mu$b DM candidate, this layer of the Earth's crust also shields DM particles.\footnote{We also considered the effect of the overburden of the Earth's atmosphere. For the DAMIC experiment at $z_{det}=$106.7 m, the Earth's atmosphere has a $\approx$27 smaller column depth ( $=\int\rho(z)dz$ ) in comparison to the Earth's crust. Due to the exponential dependence of the DM particle's final energy to the overburden column depth, the overburden from the Earth's atmosphere has a negligible effect in slowing down DM particles before reaching DAMIC's detector, for cross sections in the DAMIC exclusion region.}
\par 
In this moderately strongly interacting regime, where the DM particle mean interaction length in the Earth's crust $\lambda_{eff}$ is much smaller than the detector depth $z_{det}$, DM particles scatter off nuclei in the Earth's crust at least once during their passage. In each scattering, the DM particle loses a fraction of its initial energy. After a few scatterings in the Earth's crust, three scenarios are imaginable for each DM particle:
\begin{itemize}
	\item It scatters back to the Earth's atmosphere.
	\item Its energy falls below $E_{min}$, the minimum energy to potentially trigger the detector, before reaching the detector.
	\item It reaches the detector at the depth of $z=z_{det}$ with $E\geq E_{min}$.
\end{itemize}
\par
The minimum energy of a DM particle to scatter a silicon target nucleus in the DAMIC detector and produce the threshold recoil energy $E_r^{th}$ is
\beq
\label{eEmin}
E_{min}=\frac{(m+m_{Si})^2}{4\,m \, m_{Si}} E_r^{th}\;,
\eeq
from Equation (\ref{eEiEi-1}) for backward scattering in the CM frame, i.e. $\xi^{CM}=\pi$, and substituting for the reduced mass.
\par
The DAMIC apparatus, installed at the rather shallow underground depth of $z_{det}=$ 106.7 meters at Fermilab~\cite{Barreto2012264}, is an ideal experiment for constraining $\mu$b DM candidates. The DAMIC detector, made of silicon, accumulated a total exposure of 107 g$\cdot$days from June 2010 to May 2011, and observed a total of 106 events with an ionization signal between 40 $\rm{eV_{ee}}$ (eV electron equivalent energy) and 2 k$\rm{eV_{ee}}$.
\subsection{Monte-Carlo simulation}
We start this subsection by discussing the steps of a brute-force Monte-Carlo simulation which implements the actual path length and scattering angle distributions to find the velocity distribution of DM particles at the DAMIC detector's depth. Here are the steps of the brute-force Monte-Carlo simulation:
\par
\textit{Step 0}: For the long exposure time (from June 2010 to May 2011) of DAMIC, we considered the time-averaged initial zenith angle of DM particles to be isotropic, i.e. $cos\,\theta_0$ has a uniform distribution in $[0,1]$. (We changed the definition of $cos\,\theta$ hereafter, so that $cos\,\theta=$ 1 corresponds to a DM particle that travels along the $\hat{z}$ direction from above the detector). Also, we considered the shielding layer of the Earth's crust above the DAMIC detector as a slab of material due to the smallness of the detector depth in comparison to the Earth's radius. Therefore, only the vertical displacement of DM particles in the Earth's crust which we denote by $z$ is tracked. Knowing the zenith angle $\theta_0$, the depth $z_0$ of a DM particle at the first scattering is selected from the scattering length distribution (details below,~eq.~\eqref{eZi}). The DM particle's initial speed is sampled from the speed distribution on the Earth's surface (figure \ref{v_norm_dis}). If its initial energy is smaller than the minimum required energy to trigger the DAMIC detector, we count this DM particle as one of the particles which does not give a signal. If its energy is above the minimum energy and the particle is already at the DAMIC depth, $z_0\geq z_{det}$, it is counted as a \emph{capable} DM particle. Otherwise, the DM particle enters the first scattering iteration. 
\par
\textit{Step 1:  Choose the target nucleus.}  To calculate the energy loss, we first choose the target nucleus for the given iteration $i$ of scattering. The probability that a given DM particle scatters off a nucleus of mass number $A\in\{O,\,Si,\,Al,\,Fe\}$ is
\beq
\begin{split}
	\label{eP(A)}
	P(A)&= \frac{n_A\,\sigma_A}{\sum_A n_A\,\sigma_A}\equiv\frac{\lambda_{eff}}{\lambda_A}\;.
\end{split}
\eeq
The nuclei O (46.5$\%$), Si (28.9$\%$), Al (8.9$\%$), and Fe (4.8$\%$) make up $89.1\%$ of the Earth's crust mass constituents. We scale the mass abundances of these elements to sum up to $100\%$ to calculate the number density of each element in the Earth's crust. The relative probability $P(A)$ is a function of the DM particle's mass and the target element's mass, but it is independent of the DM-proton cross section.
\par
\textit{Step 2: Choose the CM scattering angle, $\xi_i^{CM}$, fixing the final energy.}  The scattering is isotropic in the CM frame, thus $\cos\,\xi^i_{CM}\in[-1,1]$ has a uniform distribution.   Given the target element mass number $A$, the ratio of the DM particle's energy after scattering $i$ to its energy before scattering $i$ is:
\beq
\label{eEiEi-1}
\frac{E_{i}}{E_{i-1}}=1-\frac{4\,\mu^2_{A}}{m\,m_{A}}\left(\frac{1-\cos\,\xi_i^{CM}}{2} \right)\;,
\eeq
where $\xi_i^{CM}$ is the scattering angle in the CM frame. 
\par
The scattering angle in the lab frame, $\xi_i$, and the scattering angle in the CM frame, $\xi_i^{CM}$, are related by
\beq
\label{eXi}
tan\,\xi_i=\frac{sin\,\xi_i^{CM}}{m/m_{A}+cos\,\xi_i^{CM}}\;.
\eeq

\begin{figure}[tbp]\centering
	\includegraphics[width=.65\textwidth]{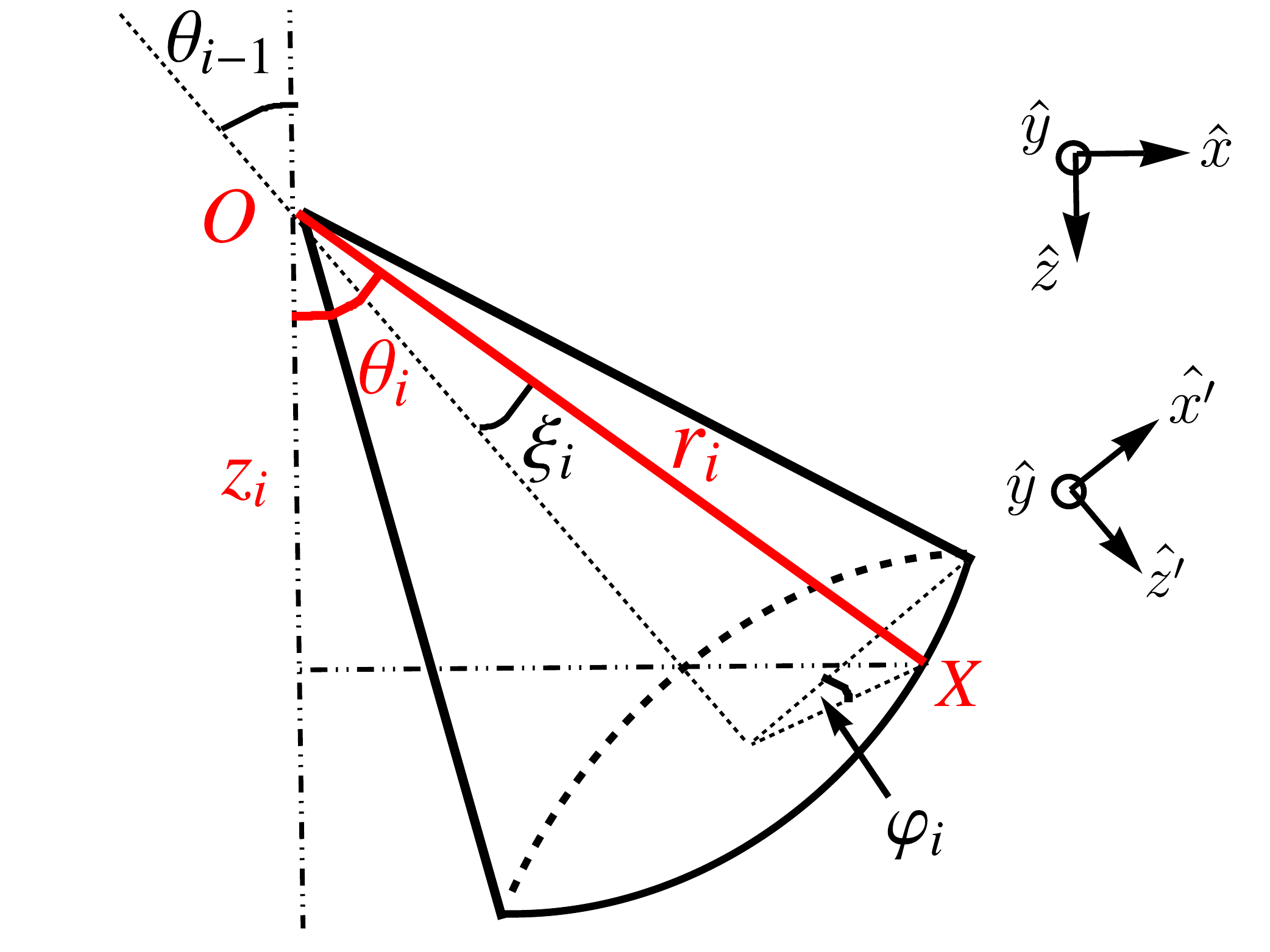}
	\caption{\label{scattering_par}The scattering parameters in the lab frame in scattering iteration $i$.}
\end{figure}
\par
\textit{Step 3: Choose the azimuthal scattering angle, $\varphi_i$, fixing the new direction.}  Figure \ref{scattering_par} shows a schematic view of the scattering parameters in the lab frame. To find the zenith angle after scattering $i$, it is sufficient to rotate $\overrightarrow{OX}$ in the primed coordinate system, $r_i(sin\,\xi_i\,cos\,\varphi_i\,,sin\,\xi_i\,sin\,\varphi_i\,,\,cos\,\xi_i)$, around the $\hat{y}$ axis with zenith angle $\theta_{i-1}$. This gives a recursive relation for the zenith angle which is also a function of angles $\xi_i$ and $\varphi_i$:
\beq
\label{eThetai}
cos\,\theta_i=\,cos\,\xi_i\,cos\,\theta_{i-1}-sin\,\xi_i\,sin\,\theta_{i-1}\,cos\,\varphi_i\;.
\eeq
\par
Generate for $\varphi_i$ a random number in $[0,2\pi]$, to calculate the zenith angle after scattering $i$ using eqs.~\eqref{eXi} and~\eqref{eThetai}.
\par
\textit{Step 4: Find the position of the next interaction.}   The probability of a DM particle scattering exactly once, off of a nucleus of specified mass number $A$ in $[r,r+dr]$, is the product of the probability that the DM particle has not scattered in the distance $r$ and the probability that it scatters in the distance $dr$ (see e.g.~\cite{2010etin.book.....T})
\beq
\begin{split}
	\label{eWA(r)}
	p_A(r)\,dr&=\frac{dr}{\lambda_A}\,e^{-r\,/\lambda_{eff}}\;.
\end{split}\eeq
The path length distribution of a DM particle passing through the Earth's crust with different constituents is the sum of all elements' probability distributions
\beq
\label{eW(r)}
p(r)=\sum_A p_A(r)=\frac{1}{\lambda_{eff}}\,e^{-r\,/\lambda_{eff}}\;.
\eeq
\par
Using eqs.~\eqref{eThetai} and~\eqref{eW(r)}, the vertical displacement between scatterings $i$ and $i+$1 can be calculated
\beq
\label{eZi}
z_i\equiv r_i\cos\,\theta_i=[ -\lambda_{eff}\,\ln\,(ran_i)]\cos \,\theta_i \;,
\eeq
where $\textit{ran}_i\in[0,1]$ is a random number representing the path length's cumulative probability distribution. 
\par
Steps 1 -- 4 are repeated until each DM particle is categorized as follows
\begin{itemize}
	\item If at any stage the depth of a DM particle is less than zero (positive direction of $\hat{z}$ taken to be downward with $z=0$ at the Earth surface), that particle is not tracked any more and is counted as one of the particles which scatters back to the Earth's atmosphere. 
	\item If the energy of a DM particle becomes smaller than the minimum energy, $E\leq E_{min}$, before reaching the DAMIC detector, that DM particle is no longer tracked and is counted as one of the events with energy below the detector's threshold.
	\item If a DM particle reaches the DAMIC detector's depth, $z\geq z_{det}$, with potentially enough energy to trigger the detector, $E\geq E_{min}$, that particle is counted as a \emph{capable} DM particle which can potentially trigger the detector. Such particles are called \emph{capable} DM particles in this paper.
\end{itemize}
\par
There is a 6-inch lead shielding layer around the DAMIC detector. Even though the thickness of this shielding layer is much smaller than the Earth's crust shielding layer of 106.7 meters, the lead shield can significantly reduce the energy of heavier DM particles for which the Earth's crust nuclei are not so well mass-matched. So, DM particles which end up in the third category go through more iterations of scattering in the lead shield. There is no need for step 1 as there is only one type of nucleus in the lead shield and $\lambda_{Pb}$ replaces $\lambda_{eff}$ in eqs.~\eqref{eW(r)} and~\eqref{eZi}. 
\subsection{Importance sampling}
Only an extremely small fraction of DM particles (as small as $\rm{8\times10^{-8}}$ for DM mass 4 GeV at the limiting cross section) reaches the DAMIC detector with enough energy to potentially produce an energy deposit above threshold (see~figure~\ref{DAMIC_att_c@sigma_MC}). This means, for a DM mass of 4 GeV, the propagation of $\rm{1.25\times10^{7}}$ particles must be simulated using the actual path length distribution to achieve one DM particle with enough energy at the DAMIC surface to potentially trigger the detector. To make the determination of the DM velocity distribution at the DAMIC detector's depth computationally feasible, we developed an importance sampling technique which reduces the computational cost of the brute-force simulation by increasing the mean path length that each DM particle travels. We account for this modification of the path length by assigning a weight factor to each simulated DM particle~\cite{method}.
\par
To increase the sample size of \emph{capable} DM particles, the path length distribution which is used in eq.~\eqref{eW(r)} is modified to
\beq
\label{eW'(r)}
q(r)=\frac{1}{(1+\delta) \lambda_{eff}}\,e^{-r\,/(1+\delta) \lambda_{eff}}\;,
\eeq
where $\delta\neq0$ shifts the average path length from $\lambda_{eff}$ to $(1+\delta)\lambda_{eff}$ while keeping the general shape of the path length distribution the same. 
\par
To correct for not sampling from the true path length distribution for the $j$-th DM particle, we assign a weight for the $i$-th scattering iteration, $w_{ij}=\frac{p(r_{i})}{q(r_{i})}$. The total weighting factor for the $j$-th particle (having $n_j$ scatterings in the Earth's crust and lead shield) is
\beq
\label{ew}
w_j=\prod_{i=0}^{n_j}w_{ij}=\prod_{i=0}^{n_j}\frac{p(r_{i})}{q(r_{i})}\;.
\eeq
This method, called importance sampling, enables our Monte-Carlo simulation to reconstruct the DM velocity distribution on the surface of the DAMIC detector using 100 to 1000 times less computational resources for $\delta=$ 0.4 -- 0.8, by wasting less computational power in simulating the events which end up losing too much energy in the Earth's crust or the lead shield. In~\cite{method}, we discuss how to calculate the distribution of any physical quantity using the importance sampling method and also validate our implementation of the strategy by comparing numerous physical distributions obtained with the brute-force and importance sampling methods.
\subsection{Cross section limits}
The DAMIC detector accumulated a total exposure of $t_e\,M_A=$ 107 g$\cdot$days from June 2010 to May 2011, and observed a total of 106 events with an ionization signal between 40 $\rm{eV_{ee}}$ and 2 k$\rm{eV_{ee}}$~\cite{Barreto2012264}. Interpreting a spectrum of ionization signal in terms of the underlying nuclear recoil energy, requires understanding the relationship between the two. For high energies, the Lindhard model~\cite{Lindhard} can be used to find the nuclear recoil energy spectrum which produces the measured ionization signal spectrum. The Lindhard model gives that the nuclear recoil energy corresponding to the maximum ionization signal 2 k$\rm{eV_{ee}}$ is 7 keV. However, for low energies the Lindhard model fails. The DAMIC team measured the ionization signal of silicon nuclei down to 60 $\rm{eV_{ee}}$, and found a significant deviation from the Lindhard model below 1 keV nuclear recoil energy~\cite{Chavarria:2016xsi}. To find the threshold nuclear recoil energy corresponding to the DAMIC electronic threshold, we extrapolate the DAMIC team's measurement of ionization factor from 60 $\rm{eV_{ee}}$ to 40 $\rm{eV_{ee}}$ and find the equivalent threshold recoil energy of $E_r^{th}=$ 550 eV. We could more conservatively limit our analysis to the ionization signal above 60 $\rm{eV_{ee}}$, where the corresponding nuclear recoil energy 700 eV has been determined experimentally. That changes the DM-proton cross-section bounds by less than one percent, but increases the minimum DM mass to which the bound applies from 1 GeV to 1.16 GeV. The functional dependence seems clear and the extrapolation small, so we adopt 40 $\rm{eV_{ee}}$ and 550 eV as the threshold. 
\par
We use the total number events observed by DAMIC, in the 40 $\rm{eV_{ee}}$ -- 2 $\rm{keV_{ee}}$ range, to find the DM-proton cross section lower bounds. Considering the observed spectral shape does not improve the cross section bounds, due to the number of DM particles with sufficient energy being exponentially dependent on the number of scatterings and hence on the cross section.
\par
For any assumed cross section, eq.~\eqref{e2_NdEr} can be integrated to calculate the expected total number of events for DAMIC's silicon target ($A=28$):
\beq\begin{split}
	\label{eDAMIC_N}
	N=&\,t_e\,M_A\,A^2\,\eta(m)\,\frac{a_c(\sigma_p)\,\rho\,\sigma_p}{4\,m\,\mu^2_p}\int_{0.55\,keV}^{7\,keV}F_A^2(E_r)\,dE_r\int_{v_{min}(E_r,A)} v\,\tilde{f}(v)\,dv\;,
\end{split}\eeq
where $\tilde{f}(v)$ is the normalized speed distribution of \emph{capable} DM particles and $\eta(m)$ is the fraction of capable DM particles at the Earth's surface. $a_c(\sigma_p)$ is the attenuation parameter (the ratio of the number of \emph{capable} DM particles at the detector to the number of capable DM particles at the Earth's surface) which is normalized to one at the Earth's surface. The extra factor of two in the denominator of eq.~\eqref{eDAMIC_N} is due the Earth-shielding of DM particles entering the Earth from below the horizon.
\par
The DM{\scriptsize ATIS} code~\cite{code}, a library written in Python and M{\scriptsize ATHEMATICA}, implemented the importance sampling Monte-Carlo simulation. For DM mass $m$ and trial DM-proton cross section $\sigma_p$, the DM{\scriptsize ATIS} code calculates the attenuation parameter and $\tilde{f}(v)$.   
\par
\begin{figure}[tbp]\centering
	\begin{subfigure}{.495\textwidth}
		\centering
		\includegraphics[width=\linewidth]{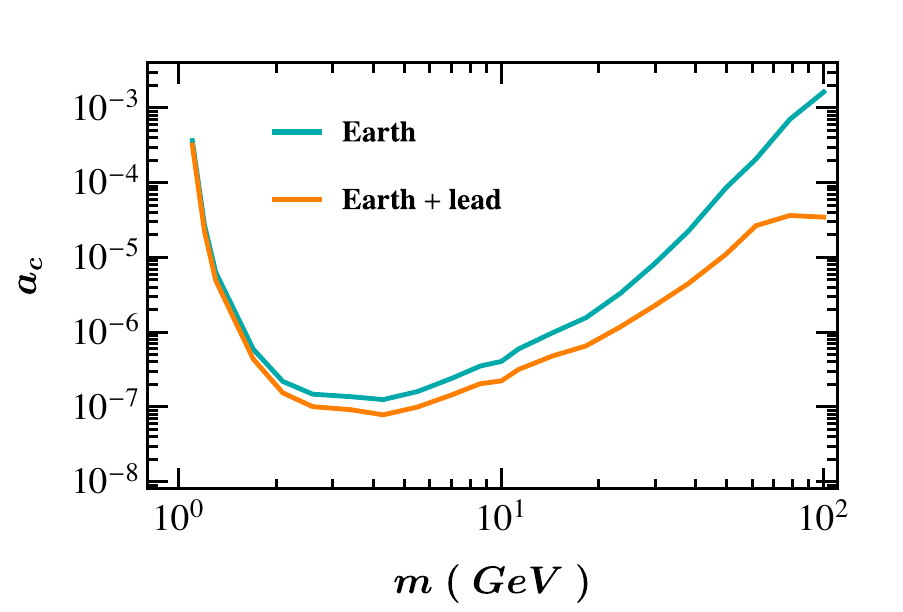}
		\caption{}
		\label{DAMIC_att_c@sigma_MC}
	\end{subfigure}\hspace{.01cm}
	\begin{subfigure}{.495\textwidth}
		\centering
		\includegraphics[width=\linewidth]{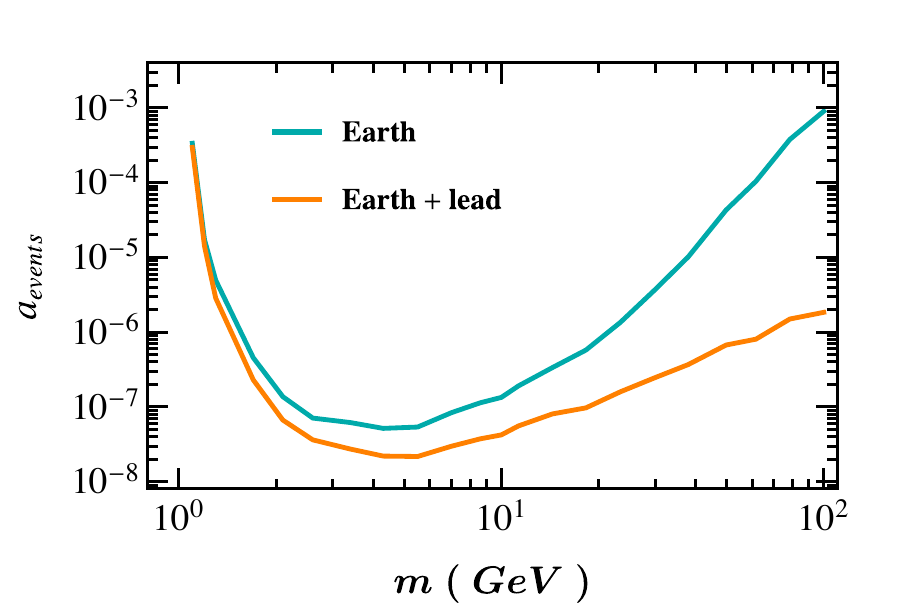}
		\caption{}
		\label{DAMIC_att_N@sigma_MC}
	\end{subfigure}
	\caption{(a) The attenuation parameter of \emph{capable} DM particles and (b) the attenuation parameter of number of events, with the DAMIC detector shielded by 106.7 meters of the Earth's crust (cyan) and also by 6 inches of the lead shield, for DAMIC's 90$\%$ CL DM-proton cross section lower bound as a function of mass that we calculate, shown in~\ref{DAMIC_bound}.}
	\label{DAMIC_att@sigma_MC}
\end{figure}
Figure~\ref{DAMIC_att_c@sigma_MC} shows $a_c$, the attenuation parameters of \emph{capable} DM particles reaching the lead shield and reaching the DAMIC detector, as a function of DM mass, corresponding to the 90\% CL DM-proton cross-section lower bounds that we calculated from DAMIC's data. 
\par
Figure~\ref{DAMIC_att_N@sigma_MC} shows the attenuation parameter for the number of events in the DAMIC energy range (the ratio of the number of events in DAMIC at its actual depth, to the number of events if the detector were on the Earth's surface), as a function of mass, corresponding to the 90\% CL cross section that we have calculated from the DAMIC data.  The difference between the Earth-only and the total attenuation parameters of the number of events increases by several orders of magnitude over the mass range considered. This demonstrates the importance of considering the thin (6 inches) lead shield around the DAMIC detector in obtaining the correct limits.
\par
\begin{figure}[h]\centering
	\includegraphics[width=\textwidth]{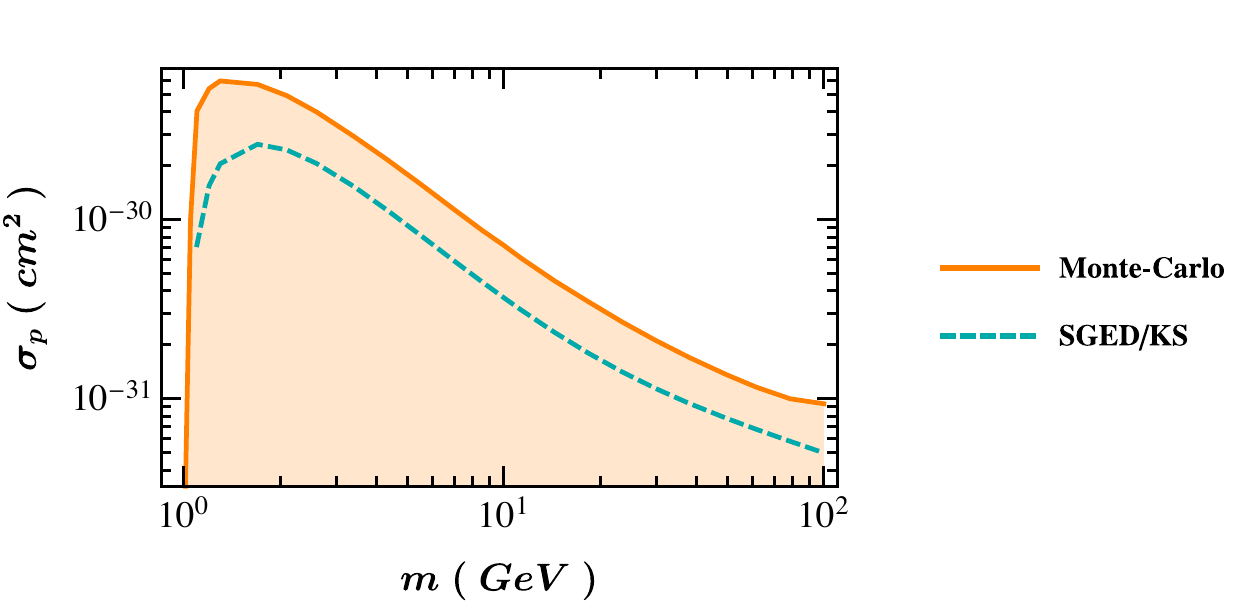}
	\caption{\label{DAMIC_bound} The 90\% CL DM-proton cross section excluded region deduced by the importance sampling Monte-Carlo simulation (solid line), for the detector depth of 106.7 meters underground. The limiting cross section obtained from the SGED\textbf{/}KS method, discussed in section~\ref{SGED}, is shown by the dashed line.}
\end{figure}

As the cross section increases, a smaller and smaller fraction of DM particles can reach the detector. As the trial cross section is decreased from some large value, the attenuation parameter increases faster than $\sigma_p^{-1}$. Therefore, the expected total number of events is a monotonically decreasing function of $\sigma_p$. The 90\% limit on the allowed (by DAMIC) cross section is defined to be the value of $\sigma_p$ for which the expected total number of events is 123. (The equivalent 90\% CL upper limit on 106 observed total number of events is 123.) The solid line in figure~\ref{DAMIC_bound} shows the 90$\%$ CL DM-proton cross section sensitivity at detector depth of 106.7 meters underground. The result from the SGED\textbf{/}KS method (dashed line), which we discuss briefly in the next subsection and in detail in~\cite{method}, is added for comparison.
\par
\begin{figure}[h]\centering
	\includegraphics[width=.9\textwidth]{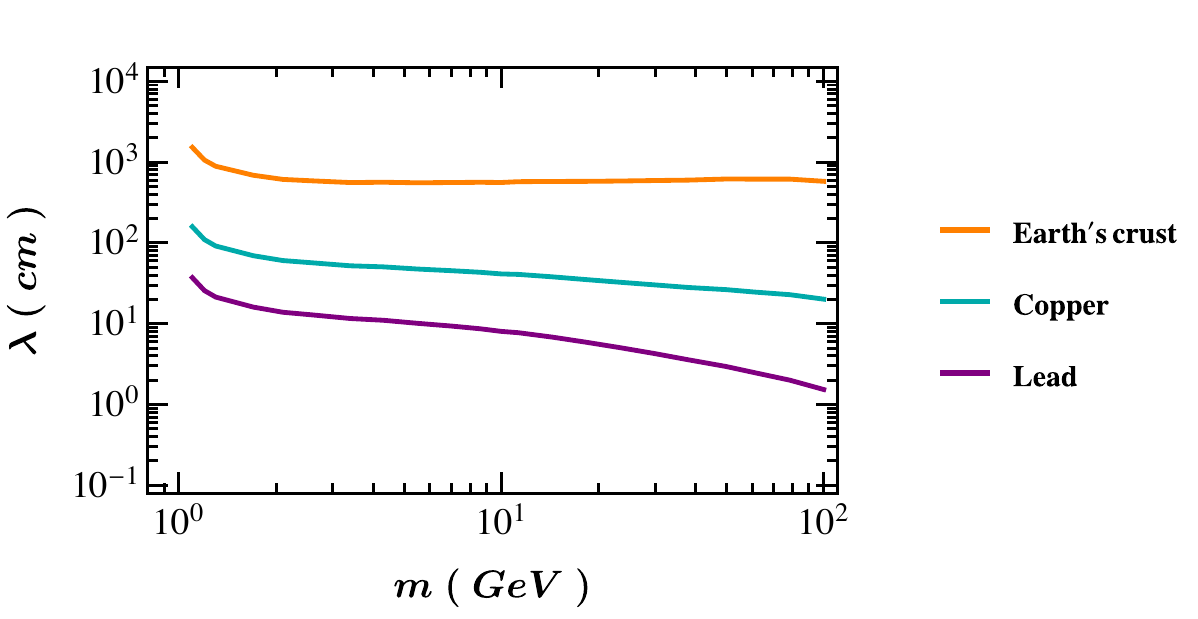}
	\caption{\label{DAMIC_lambdas}The mean interaction length of DM particles in the 106.7 meters of the Earth's crust (orange), in the 0.5 cm of the copper shield around the DAMIC detector (cyan), and in the 6 inch ($\simeq$ 15 cm) lead shield around the DAMIC detector (purple line) using the 90\% CL DM-proton cross section lower bounds from the DAMIC detector.}
\end{figure}
Using our 90\% CL cross section limits as a function of mass that we have calculated from the DAMIC data, the mean interaction lengths in the Earth's crust, in a target of copper, and in a target of lead are given in figure~\ref{DAMIC_lambdas}. The mean interaction lengths in a target of copper for these cross sections are at least ten times bigger than the copper shield thickness ($0.5$ cm) around the DAMIC detector. This justifies our shielding model which ignores the copper shield around the DAMIC detector.
\subsection{Properties of DM particles that successfully triggered DAMIC}\label{dist}
It is enlightening to contemplate some properties of the DM particles that produce events in the DAMIC detector, for representative DM masses 1.7, 10, and 100 GeV.  In each case, we use the 90\% CL sensitivity-limit cross section for these DM masses, i.e. $\sigma_p = 5.7 \times\rm{10^{-30}}$, 7.2$\times\rm{10^{-31}}$, and 9.3$\times\rm{10^{-32}\,cm^2}$.
\par
\begin{figure}[tbp]\centering
	\begin{subfigure}{.495\textwidth}
		\centering
		\includegraphics[width=\linewidth]{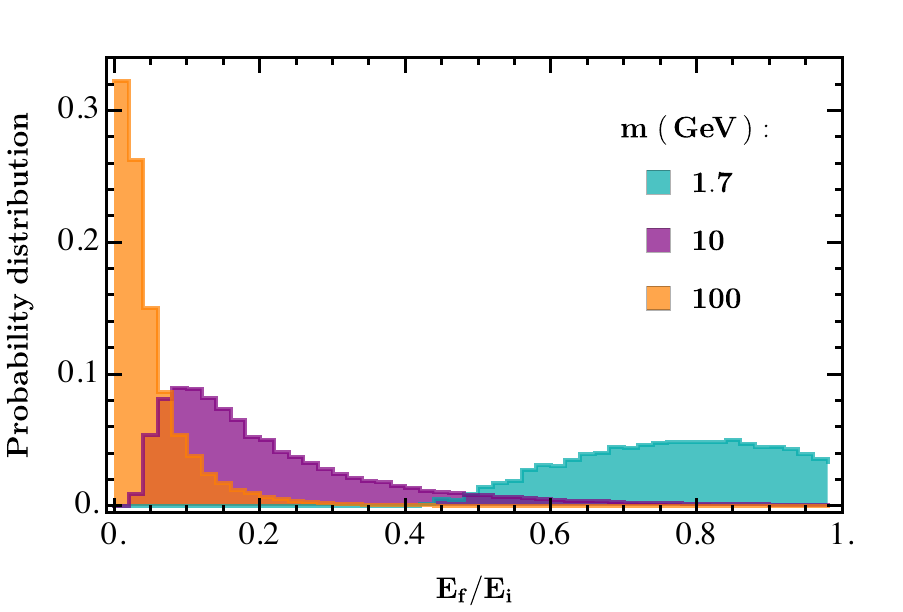}
		\caption{}
		\label{EfEi_mD}
	\end{subfigure}\hspace{0.01cm}
	\begin{subfigure}{.495\textwidth}
		\centering
		\includegraphics[width=\linewidth]{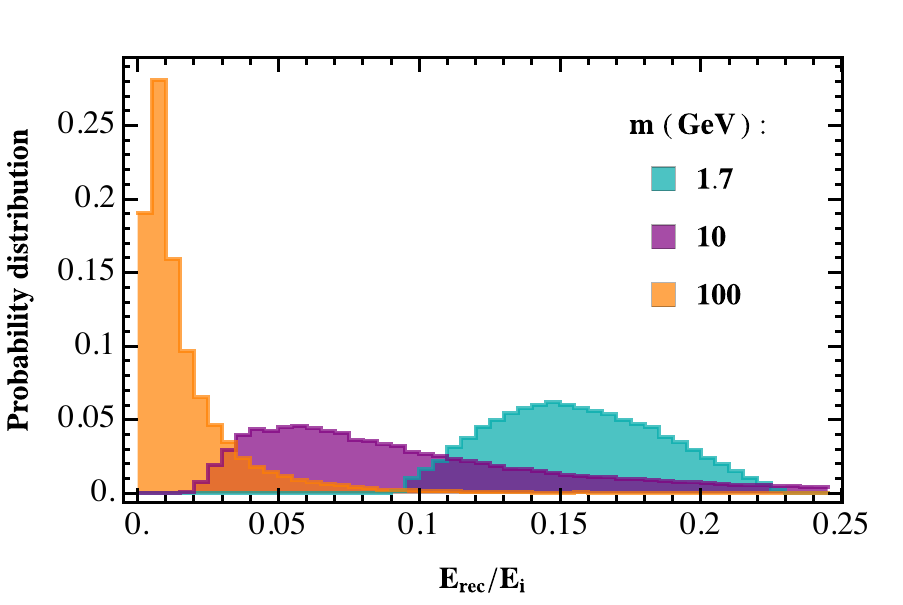}
		\caption{}
		\label{Er_mD}
	\end{subfigure}
	\caption{The distributions of the ratios of (a) the energy after penetrating the overburden and (b) the nuclear recoil energy, to the initial energy, for \emph{successful} DM particles with masses of 1.7, 10, 100 GeV using their corresponding DM-proton cross section lower bounds.}
	\label{EmD}
\end{figure}

Figure~\ref{EmD} shows the distributions of the final energies and the recoil energies of \emph{successful} DM particles, i.e. DM particles that reach the DAMIC detector and scatter the detector giving a deposit-energy above the DAMIC threshold, $E_r=$ 550 eV. In this mass range, as the DM mass increases, DM particles can lose a larger fraction of their initial energy while passing through the Earth's crust and the lead shield, and deposit a smaller fraction of their initial energy to the silicon nuclei of the DAMIC detector, and still produce an above threshold energy-deposit. 
\par
\begin{figure}[tbp]\centering
	\begin{subfigure}{.33\textwidth}
		\centering
		\includegraphics[width=\linewidth]{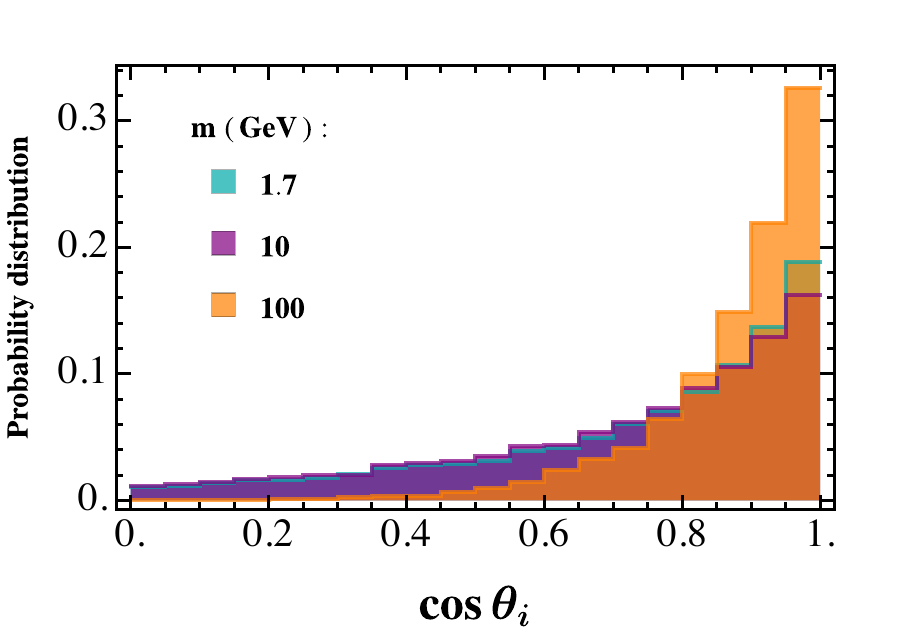}
		\caption{}
		\label{thetaE_mD}
	\end{subfigure}\hspace{0.001cm}
	\begin{subfigure}{.328\textwidth}
		\centering
		\includegraphics[width=\linewidth]{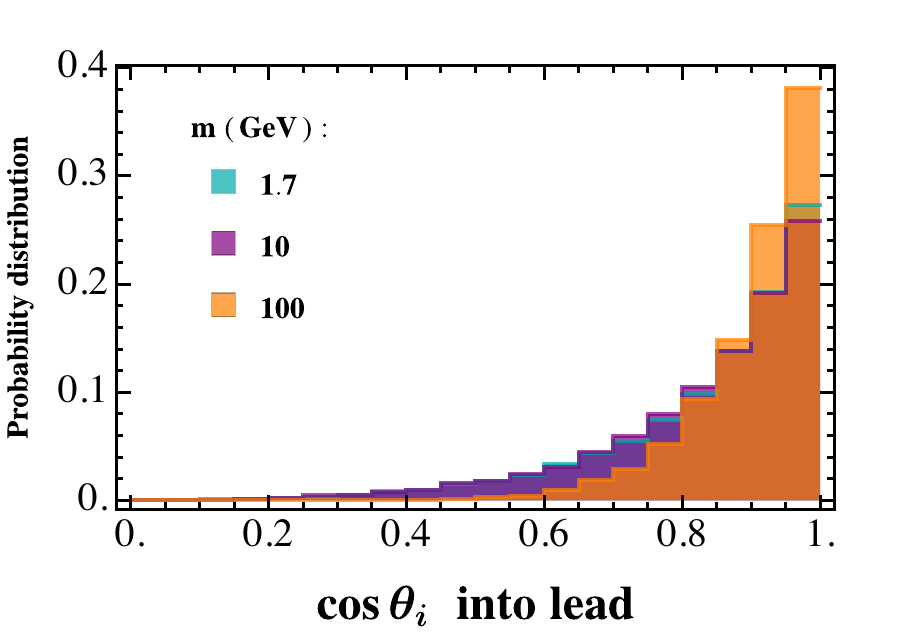}
		\caption{}
		\label{thetaPb_mD}
	\end{subfigure}\hspace{0.001cm}
	\begin{subfigure}{.327\textwidth}
		\centering
		\includegraphics[width=\linewidth]{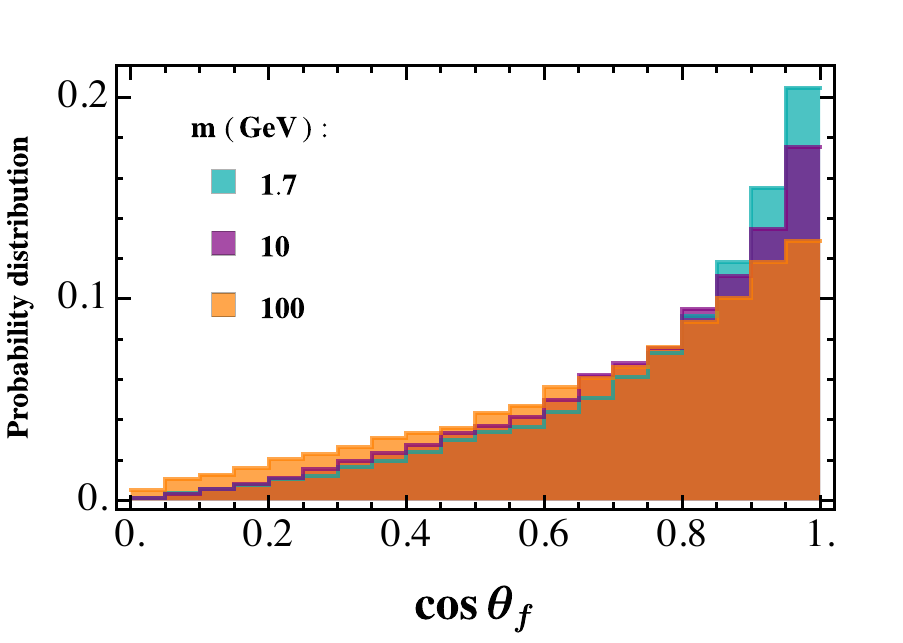}
		\caption{}
		\label{thetaSi_mD}
	\end{subfigure}
	\caption{The zenith angle distributions of \emph{successful} DM particles (a) on the Earth's surface, (b) on the lead shield, and (c) on the DAMIC detector with masses of 1.7, 10, 100 GeV and their corresponding DM-proton cross section lower bounds.}
	\label{theta_mD}
\end{figure}
\begin{figure}[tbp]\centering
	\begin{subfigure}{.495\textwidth}
		\centering
		\includegraphics[width=\linewidth]{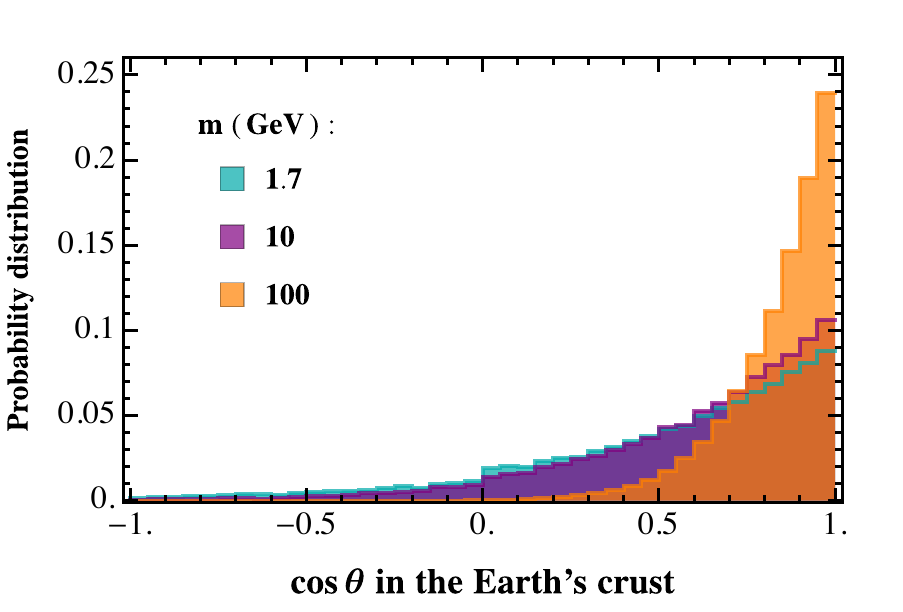}
		\caption{}
		\label{thetaE_D}
	\end{subfigure}\hspace{0.01cm}
	\begin{subfigure}{.495\textwidth}
		\centering
		\includegraphics[width=\linewidth]{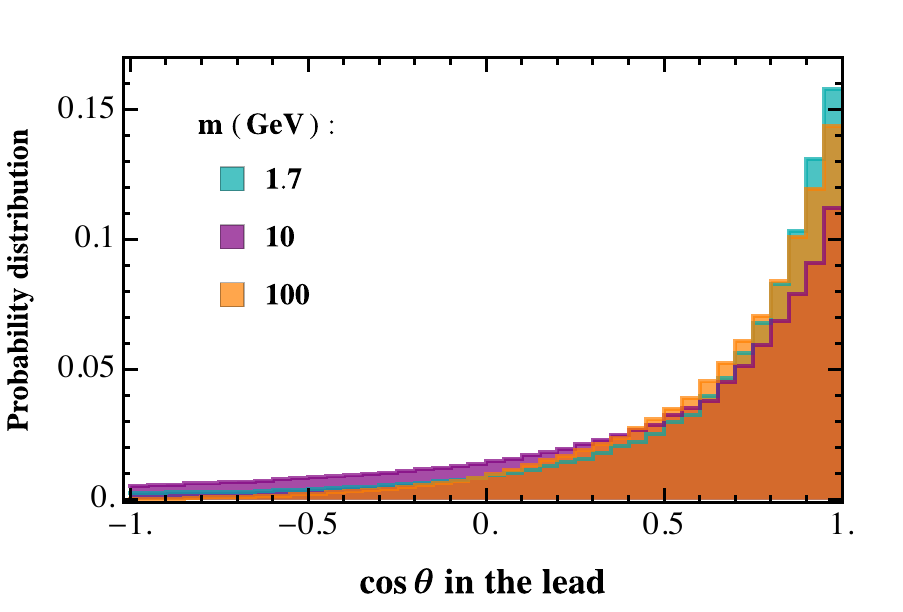}
		\caption{}
		\label{thetaPb_D}
	\end{subfigure}
	\caption{The distributions of the zenith angles of all scatterings in the trajectory of all \emph{successful} DM particles for masses 1.7, 10, and 100 GeV when passing through (a) the Earth's crust and (b) the lead shield.}
	\label{thetaD}
\end{figure}

Figure~\ref{theta_mD} shows the zenith angle distribution of successful DM particles. Although successful DM particles mainly enter and exit the Earth's crust vertically for any mass, the heavier DM particles exit the lead shield with a much broader zenith angle distribution. This broadening is due to the similarity of the DM and target nuclei masses in the Earth's crust and the lead shield.  
Figures~\ref{thetaE_D} and \ref{thetaPb_D} show distributions of the zenith angles in all scatterings of successful DM particles in the Earth's crust and the lead shield. Depending on the mass of DM particles and the material that these DM particles pass through, the distribution of the averaged zenith angle changes. Clearly, DM particles that trigger the detector successfully do not move vertically with zero deflection angle.   
\subsection{Deficiencies of the  SGED\textbf{/}KS method }\label{SGED}
Starkman, Gould, Esmailzadeh and Dimopolous~\cite{Starkman} proposed a continuous energy-loss and vertical propagation approximation, to estimate the maximum cross section for which a given experiment is sensitive to strongly interacting DM particles.   They took the inverse energy-loss-length, $\frac{d E}{E dz}$ to be $ \langle f \rangle / \lambda$ where $\langle f \rangle$ is the mean fractional energy loss for isotropic scattering and $\lambda$ is the scattering length in the material.  This estimation method was applied by Kouvaris and Shoemaker~\cite{Kouvaris2014},  to analyze DAMIC's data.  
\par
\begin{figure}[tbp]\centering
	\begin{subfigure}{.495\textwidth}
		\centering
		\includegraphics[width=\linewidth]{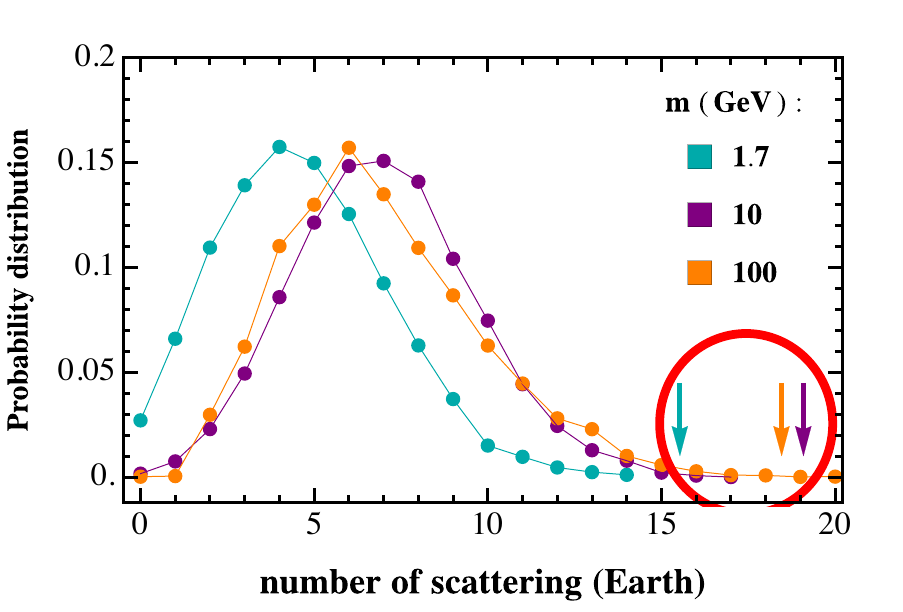}
		\caption{}
		\label{nE_mD}
	\end{subfigure}\hspace{.01cm}
	\begin{subfigure}{.495\textwidth}
		\centering
		\includegraphics[width=\linewidth]{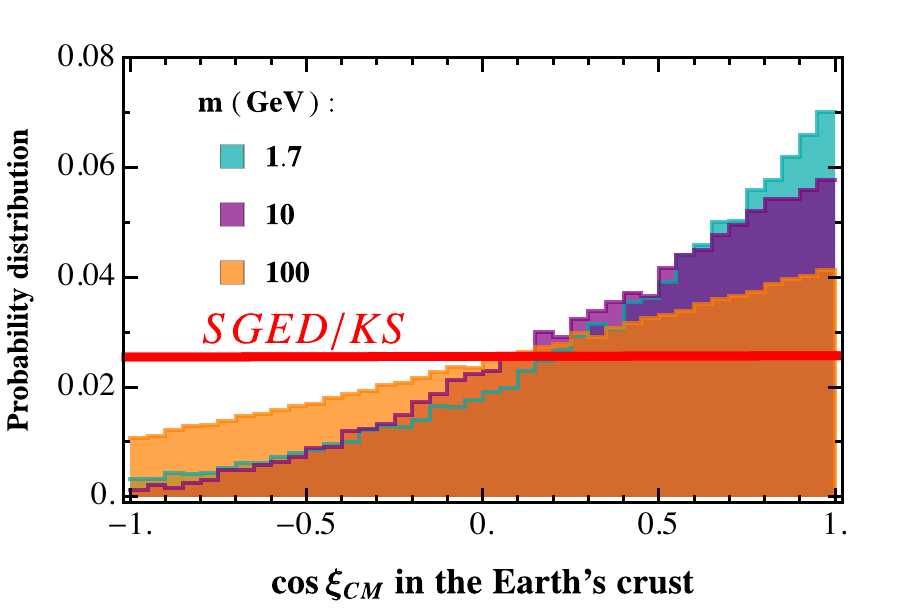}
		\caption{}
		\label{avexiE_D}
	\end{subfigure}
	\caption{The distribution of (a) the number of scatterings and (b) the scattering angle in the CM frame in the all scatterings for \emph{successful} DM particles with masses 1.7, 10, and 100 GeV. Successful DM particles exploit the tail of (a) the path length distribution and (b) the CM frame scattering angle distribution to lose less energy. The arrows indicate the number of scatterings averaged over all trajectories (including unsuccessful ones) for particles propagating a distance $z_{det}$, i.e. $\frac{z_{det}}{\lambda_{eff}}$ in the Earth's crust and $\frac{z_{Pb}}{\lambda_{Pb}}$ in the lead shield. These are the values which are used in the SGED\textbf{/}KS approximation which explains the failure of this approximation.}
	\label{SGEDd}
\end{figure}

Figure~\ref{DAMIC_bound} shows that the SGED\textbf{/}KS method underestimates the true sensitivity range of DAMIC. Figures~\ref{nE_mD} and~\ref{avexiE_D} show that DM particles that trigger the detector are exploiting the tails of the path length and scattering angle distributions to lose less energy. Using mean values of these quantities is very inaccurate. The problem is most acute when the DM energy-loss per interaction can be significant due to the DM and target nucleus masses being comparable. Figures~\ref{theta_mD} and ~\ref{thetaD} show that the assumption of purely vertical propagation is inaccurate.  For a more detailed discussion of the problems with the SGED\textbf{/}KS method see~\cite{method}. 
\par
\begin{figure}[tbp]\centering
	\includegraphics[width=\textwidth]{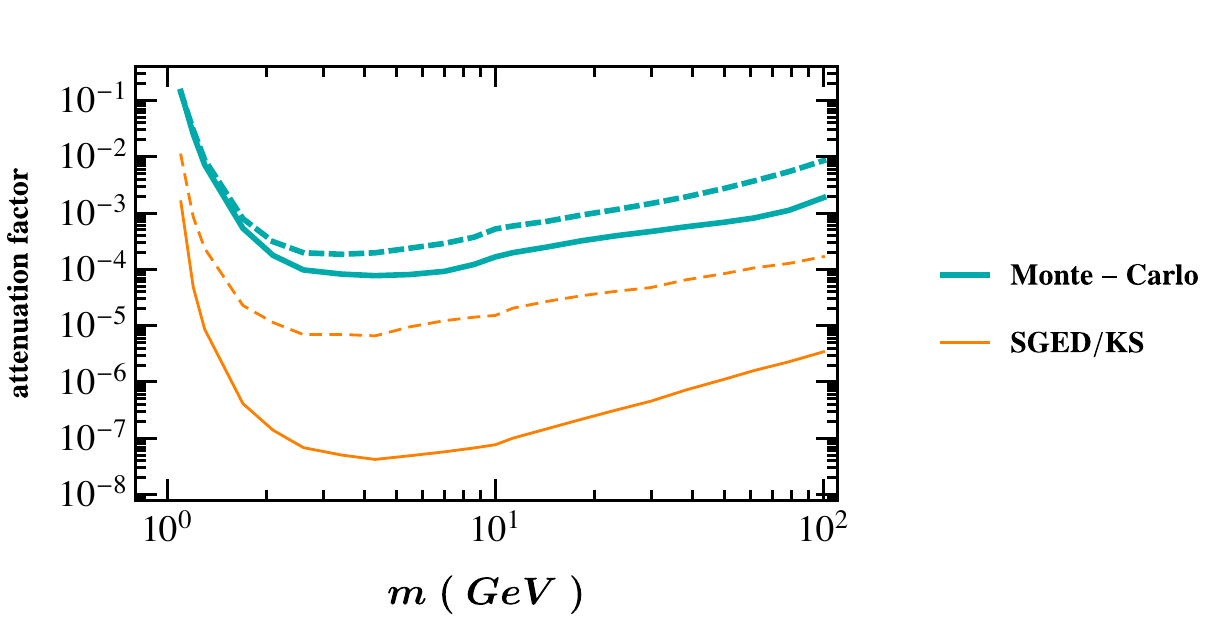}
	\caption{\label{DAMIC_att@sigma_SGED}The attenuation parameters for (solid) total number of detected events and (dashed) number of ``capable'' events, calculated with the Monte-Carlo (cyan, solid) and SGED\textbf{/}KS (orange, thin solid) methods.}
\end{figure}

Although our lower bounds on the DM-proton cross section are larger by only a factor of 1.8 -- 5.6 than KS, this difference corresponds to a four order-of-magnitude larger expected total number of events for the KS limiting cross section (compare orange-solid and cyan-solid lines in figure~\ref{DAMIC_att@sigma_SGED}). This reveals how badly the SGED\textbf{/}KS method fails to calculate the energy loss of DM particles propagating in the Earth and the lead shield. Nonetheless, as explained in~\cite{method}, the SGED approximation work to get an order-of -magnitude estimate of the cross section sensitivity which was the original SGED purpose; the problem is to apply it for extracting a limit as done in KS~\cite{Kouvaris2014}.
\section{Conclusion}
In this work, we have derived the constraints on moderately interacting dark matter which follow from the XQC and DAMIC experiments. Previous analyses made short-cuts which significantly reduced the apparent constraining-power of these experiments and indicated a window for $\sigma_p \sim$ 0.1 -- 1 $\mu$b, for dark matter mass in the 1 -- 100 GeV range.  Our new, more stringent limits, close the window for 1 -- 8 GeV and exclude the possibility that Dark Matter having a mass 1 -- 100 GeV, has moderate interactions with nucleons, when combined with limits from other experiments.  
\par
Our upper limit on the DM-nucleon scattering cross section from the XQC sounding-rocket experiment~\cite{McCammon2002} is a factor 8.9 lower (more restrictive) than reported by Erickcek et al.~\cite{Erickcek2007}; where analysis made the assumption that the rocket body shields the detector which is not the case in the the relevant cross section range.  Our results refine but generally confirm, the earlier analysis of XQC of~\cite{ZF_window}.   
\par
The DAMIC experiment operating 106.7m underground, is shielded by the Earth and lead shielding.  Due to energy loss in the overburden, only a tiny fraction of the flux of DM particles at the surface of the Earth, has sufficient energy to trigger the detector.  This means that there is a limiting cross section, above which DAMIC is insensitive to DM.  In order to determine what cross-sections DAMIC can exclude, one must accurately determine the attenuation of DM particles \emph{capable} of producing an above threshold energy-deposit, as a function of cross section. The analysis of Kouvaris and Shoemaker~\cite{Kouvaris2014}, used a poor approximation to the energy loss and \textit{overestimated} the attenuation enormously (see figure~\ref{DAMIC_att@sigma_SGED}).  With the limiting cross section we find, the KS approximation implies 100\% attenuation, and for their reported limit, they overestimate the attenuation by a factor $\approx 10^4$.  With a correct treatment of energy loss, we find that DAMIC is sensitive to cross sections a factor 2 or more higher than KS reports, extending the exclusion range of DAMIC to higher cross sections and overlapping the new XQC upper bound.    
\par
Besides the exclusion of this type of Dark Matter, our results shed light on the process of energy loss in the Earth or another overburden.  We showed that the commonly-used SGED~\cite{Starkman} approximation, based on the mean number of interactions corresponding to an assumed cross section and vertical propagation to the detector, and the average energy-loss per interaction, is grossly misleading in some circumstances. The problem is particularly severe when the mass of nuclei in the overburden is within a factor few-to-10 of that of the DM particle, so that a substantial fraction of the DM's energy can be lost in each collision. In this case, there is a population of trajectories with lower-than-average scattering angles and longer-than-average path lengths between interactions, which can arrive at the detector with sufficient energy to be capable of triggering the detector. Such a population is completely missed in the SGED approximation used by \cite{Kouvaris2014} to analyze the DAMIC experiment, resulting in a four order-of-magnitude larger number of events expected in DAMIC over a large mass range (see figure~\ref{DAMIC_att@sigma_SGED}).  
We present more details of this phenomenon in a companion paper~\cite{method}, where we also present an importance sampling technique to reduce computational time by a factor 100 -- 1000.  This makes the Monte-Carlo simulation of trajectories and energy loss feasible, even when the attenuation parameter is smaller than $10^{-7}$.  The DM{\scriptsize ATIS} code, which we developed for this study, will be publicly available on-line~\cite{code}. 
\par
\begin{figure}[tbp]\centering
	\includegraphics[width=.9\textwidth]{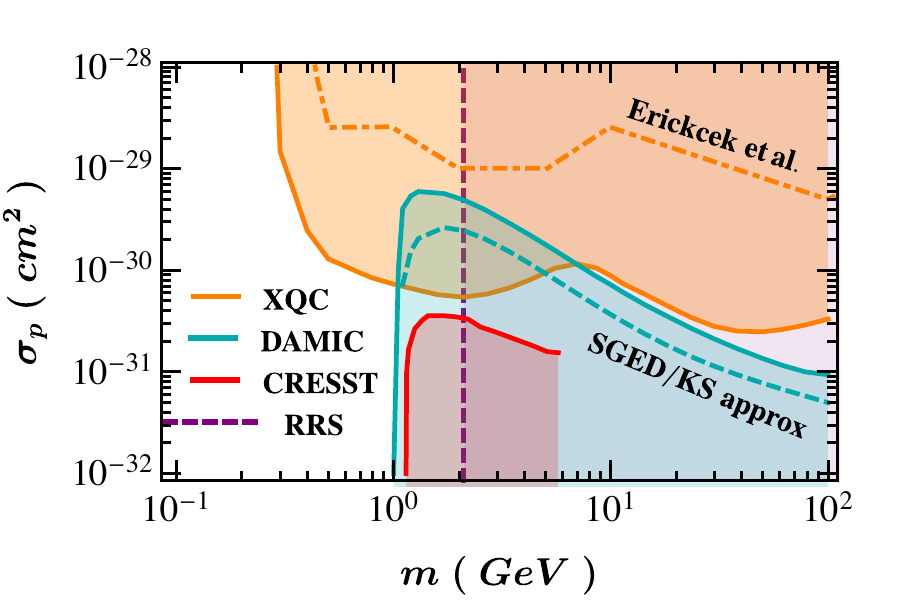}
	\caption{\label{XQC&DAMIC}The 90\% CL DM-proton lower (cyan lines) and upper (orange lines) bound cross sections of a $\mu$b DM candidates using the DAMIC and XQC observed spectra. Limits from CRESST (red line, taken from~\cite{ZF_window}), which were the strongest lower bounds before DAMIC, are shown for comparison. The limits from RSS (purple line, taken from~\cite{RICH1987}) balloon experiment exclude a $\mu$b DM candidate heavier than 2.1 GeV.}
\end{figure}

Figure~\ref{XQC&DAMIC} summarizes our 90\% CL DM-proton spin-independent cross section bounds for DM masses of 0.3 -- 100 GeV, using  the XQC and DAMIC energy-deposit spectra. This closes the previously-existing window for $\mathcal{O}(\mu$b) cross sections for DM masses in the 1 -- 8 GeV range.  All of the limits presented here assume the standard DM velocity distribution.  For completeness, we performed simulations assuming other Maxwellian velocity distributions;  the results are presented in Figure \ref{XQC-vpeak}.

\noindent {\bf Acknowledgements:}  We thank A. Erickcek and D. McCammon for information about XQC and the analysis of~\cite{Erickcek2007}. MSM acknowledges support from James Arthur Graduate Assistantship;  the research of GRF has been supported by NSF-PHY-1212538 and AST-1517319.

\bibliography{g1}
\end{document}